\documentclass[useAMS,usenatbib,usegraphicx]{mn2e}

\usepackage{graphicx}
\usepackage{amssymb}

\newcommand{\apjs}{ApJS}
\newcommand{\apj}{ApJ}
\newcommand{\apjl}{ApJL}

\newcommand{\mnras}{MNRAS}

\newcommand{\nar}{New Astron. Rev.}
\newcommand{\msun}{\mbox{M}_\odot}
\newcommand{\sigmap}{\sigma_{p}}
\newcommand{\sigmab}{\sigma_{B}}
\newcommand{\meanp}{\langle{p_0}\rangle}
\newcommand{\meanb}{\langle{B}\rangle}
\newcommand{\meanx}{\langle{x}\rangle}
\newcommand{\etam}{\eta_0}

\newcommand {\beq}{\begin {eqnarray}}
\newcommand {\eeq}{\end {eqnarray}}
\newcommand{\ergs}{\,\rm{erg}\,\,\rm{s}^{-1}}

\begin{document}

\title[Young pulsars as ultra-luminous X-ray sources]
{Young rotation-powered pulsars as ultra-luminous X-ray sources}

\author[A. S. Medvedev \& J. Poutanen]{Aleksei S. Medvedev$^{1,2}$ and Juri Poutanen$^1$\thanks{E-mail:
juri.poutanen@oulu.fi}\\
$^{1}$Astronomy Division, Department of Physics, P.O. Box 3000, FI-90014 University of Oulu, Finland\\
$^{2}$Finnish Centre for Astronomy with ESO (FINCA), University of Turku, V\"{a}is\"{a}l\"{a}ntie 20, 21500 Piikki\"{o}, Finland
}

\pagerange{\pageref{firstpage}--\pageref{lastpage}}
\pubyear{2013}

\date{Accepted 2013 February 24.  Received 2013 February 20; in original form 2013 January 16}

\maketitle
\label{firstpage} 

\begin{abstract}
The aim of the present paper is to investigate a possible contribution of the rotation-powered pulsars  
and pulsar wind nebulae to the  population of ultraluminous X-ray sources (ULXs). 
We first develop an analytical model for the evolution of the distribution function of pulsars over the spin period and 
find both the steady-state  and the time-dependent solutions.   
Using the recent results on the X-ray efficiency dependence on pulsar characteristic age, we then 
compute the  X-ray luminosity function  (XLF) of rotation-powered pulsars. 
In a general case it has a broken power-law shape with a high luminosity cutoff, 
which depends on the distributions of the birth spin period and the magnetic field.  

Using the observed XLF  of sources in the nearby galaxies and the condition that 
the pulsar XLF does not exceed that, we find the allowed region for the parameters describing the birth period distribution. 
We find that the mean pulsar period should be greater than 10--40~ms. 
These results are consistent with the constraints obtained from the X-ray luminosity of core-collapse supernovae. 
We estimate that the contribution of the rotation-powered pulsars to the ULX population is at a level exceeding 3 per cent. 
For a wide birth period distribution, this fraction grows with  luminosity  and above $10^{40}\ergs$ pulsars can dominate the ULX population. 
\end{abstract}


\section{Introduction}

Ultraluminous X-ray sources (ULXs) are non-nuclear, point-like objects with apparent X-ray luminosity exceeding the Eddington limit 
for a stellar mass black hole \citep[see][ for a review]{feng_soria11}. 
These objects were  discovered by the {\it Einstein} satellite in nearby star-forming galaxies \citep{long83,fab88,fab89,fabtri97,stocke91}. 
Observations with  {\it Chandra} and {\it XMM-Newton} satellites have extended the  sample of probable ULXs to 
about 500 sources \citep{Swartz2011,walton11}.

There are several hypotheses about the nature of ULXs. The most popular models at this moment involve stellar-mass objects similar 
to SS~433 with the supercritical regime of accretion and mild beaming with beaming factor $1/b = 4\pi/\Omega \lesssim 10$ 
\citep{K01,fabrika04,PLF07}, 
or the accreting intermediate mass black holes (IMBH) with masses $M\sim$10$^{3}$--10$^{5}\,\msun$ \citep[e.g.][]{CM99}. 

Many ULXs show spectral variability \citep{Kajava09} typical for  the accreting black holes. 
The presence of soft thermal excesses sometimes seen in the ULX spectra \citep{KCPZ03,MF03} can be used as an argument of a 
large emission region size, which is either a signature of an IMBH or, alternatively,  
a large extended photosphere in a strong outflow from stellar-mass objects accreting at super-Eddington rates \citep{PLF07}. 
The best IMBH candidates, the brightest ULXs, M82~{X-1} and ESO 243--49~{HLX-1},  
show  spectral states similar to those seen in Galactic sources \citep{GBW09,FK10,SFL11},  but at higher luminosities. 
However,  IMBHs cannot dominate the ULX population, 
because many ULXs are  associated with the star-forming regions \citep{Swartz2009} and young stellar clusters,
but are clearly displaced from them by 100--300 pc \citep{Zezas02b,Kaaret04,Ptak2006,PFVSG12}. This in turn 
strongly argues in favour of the young, massive X-ray binaries as the ULX hosts that have been 
ejected from the stellar clusters by gravitational interactions during cluster formation and/or due to the supernova (SN) explosions. 
It is very likely, however, that the ULX class is not homogeneous, but contains different kinds of objects. 

For example, some of the bright, steady ULXs could be young, luminous rotation-powered pulsars.
Earlier studies \citep{sewart88,becker97} 
suggest that X-ray luminosity of the pulsars is correlated with the rotation energy losses $L= \eta \dot E_{\rm rot}$. 
The efficiency $\eta$, which defines the amount of rotational energy losses converted to the X-ray radiation, 
was found to be nearly constant. 
Later,  using a more complete sample of X-ray rotation-powered pulsars, \citet{possenti02}
showed that the X-ray luminosity  depends on the rotational energy loss as a power law $L\propto \dot E_{\rm rot}^{1.34}$.
 
\citet{perna04}  performed first Monte Carlo simulations of the X-ray luminosity function (XLF) of rotation-powered pulsars. 
In order to describe the luminosity evolution of the pulsars together with the evolution 
of the spin period due to the magnetic-dipole radiation losses, they used the efficiency -- characteristic age 
dependence from \citet{possenti02}. 
They considered the distribution functions of  pulsars over the magnetic field and the birth spin period given by 
\citet{arz02} and showed that rotation-powered pulsars can be very bright X-ray sources with luminosities 
$L> 10^{39}\ergs$. 

Recent investigation of the X-ray properties of rotational-powered pulsars conducted by \citet{vink11} 
revealed a more complicated efficiency--age dependence. 
Using the new  data from  {\it Chandra} observatory \citep{kargaltsev08}, they find that radiative efficiency is not constant for pulsars with age $< 1.7\times10^4$~yr, but depends on the characteristic age. 
These new results may strongly affect the XLF, increasing the number of the most luminous pulsars. 

In the present paper we develop a model for the XLF of the rotation-powered pulsars, 
taking into account the recently discovered efficiency--age dependence. 
In Section~\ref{s:model} we present the analytical model describing the evolution of the pulsar periods 
and find both the steady-state and the time-dependent solutions. 
Section~\ref{s:obser} is devoted to the observational constraints on the model 
parameters for the birth period and magnetic field distribution that can be obtained 
from the core-collapsed SNe and the observed XLF of the sources in the nearby galaxies. 
In Section~\ref{s:ULX} we obtain the birth period and magnetic field distributions for the brightest  pulsars and 
estimate the possible contribution of young pulsars to the ULX population. 
We summarize in Section~\ref{s:summary}.

\section{Model}
\label{s:model}

\subsection{Basic equations}
\label{sec:basic}

A pulsar is described by two parameters: its birth period $p_0$ and the magnetic field $B$, which is assumed to
be constant over its lifetime.  
We consider lognormal distributions for both $B$  and $p_0$, 
with the probability density for the decimal logarithm $\log x$ in the following form: 
\begin{equation}
G (\log x; \log \meanx, \sigma_x) = \frac{1}{\sqrt{2\pi} \ \sigma_x} e^{-\displaystyle \frac{\log^2 (x/\meanx)}{2\sigma_x^2}} . 
\end{equation}
The mean and the standard deviation (scale) for the two distributions are 
($\log\meanb$, $\sigmab$) and ($\log\meanp$, $\sigmap$). 

\begin{figure}
\centering
\includegraphics[width=0.9\linewidth]{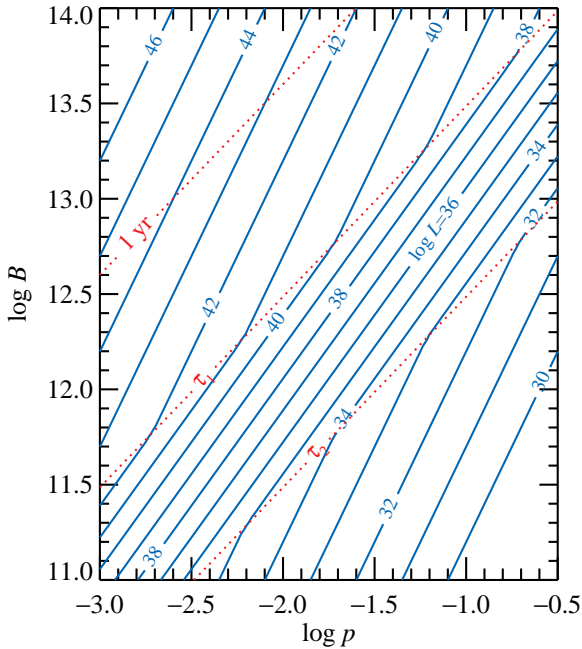}
\caption{Contour plots of constant X-ray luminosity $\log L$ (solid lines) 
and the characteristic age $\tau_{\rm c}$ (dotted lines)
on the plane magnetic field -- period for $\etam=1$. 
}
\label{f:lum_tau}
\end{figure}

The pulsar period at a given age is calculated using a simple model, where the rotational energy losses are dominated 
by the magnetic dipole radiation \citep[see e.g.][]{gosh07}:  
  \begin{equation}
        \dot E_{\rm rot} = -  I \Omega \dot \Omega =  \frac{2R^6}{3 c^3} B^2 \Omega^4   ,  
  \end{equation}  
 where $\Omega=2\pi/p$ is the pulsar rotational frequency,  $R$ is the neutron star radius and $I$
 is its moment of inertia. 
 We ignore the factor depending on the angle between the dipole and the rotational axis 
 to be consistent with previous studies. 
The evolution of the pulsar period and the frequency are  described by equations
 \begin{equation}
        p\ \dot p = \frac{1}{2}\alpha B^2, \quad         \dot \Omega = - \frac{\alpha}{8\pi^2} B^2 \Omega^{3} . 
  \end{equation}
where 
 \begin{equation}
  \alpha = \frac{16 \pi^2}{3} \frac{R^6}{I c^3}.
 \end{equation}  
The time-dependence of the pulsar period is then    
 \begin{equation}\label{eq:p_t}
        p(t) = \sqrt{p_0^2 + \alpha B^2 t },
  \end{equation} 
with the characteristic spindown age 
 \begin{equation} \label{eq:char_age}
\tau_{\rm c}=\frac{p}{2\dot p} = t + \frac{p_0^2}{\alpha B^2} = 
5\times 10^{14} p^2 B_{12}^{-2}  \ \mbox{s} , 
  \end{equation} 
where we assumed $I_{45}=1$ and $R_6=1$ (i.e. $\alpha \approx 2\times 10^{-39}$ cgs) and 
used standard notations $Q=10^x Q_x$ in cgs units.   
On the $B$--$p$ plane, using equation (\ref{eq:char_age}) 
we can identify the lines  of constant characteristic age 
$B_{12}=3.9\ p_{-3}\ \tau_{\rm c, yr}^{-1/2}$ (see dotted lines in Fig.~\ref{f:lum_tau}).

We estimate the X-ray  luminosity $L$ of a pulsar (and a pulsar wind nebula, PWN) 
from its period and period derivative following \citet{vink11}.   
For simplicity, we approximate the efficiency--age dependence with a simple relation: 
 \begin{equation}\label{eq:xrayeff}
\eta=L/  \dot E_{\rm rot}= \left\{ \begin{array}{ll} 
\etam , & \mbox{if}  \  \tau_{\rm c} \leqslant \tau_1 , \\
\etam \ (\tau_1/\tau_{\rm c})^2 , & \mbox{if}\ \tau_1 \leqslant  \tau_{\rm c} \leqslant \tau_2 , \\
10^{-4} , & \mbox{if} \  \tau_{\rm c} \geqslant \tau_2 , 
\end{array} \right.  
\end{equation}  
where $\tau_1=170\,\mbox{yr}\times \etam^{-1/2}$ and $\tau_2=1.7\times 10^4$ yr. 
This is equivalent to 
  \begin{equation}\label{eq:luminos}
L= \left\{  \begin{array}{ll}
4\times 10^{31} p^{-4}  B_{12}^{2} \etam \ \ergs , & \mbox{if}  \  \tau_{\rm c} \leqslant \tau_1 , \\
 4\times 10^{21} p^{-8}  B_{12}^{6}  \ \ergs, & \mbox{if}\ \tau_1 \leqslant  \tau_{\rm c} \leqslant \tau_2 , \\
 4\times 10^{27} p^{-4}  B_{12}^{2}  \ \ergs, & \mbox{if} \  \tau_{\rm c} \geqslant \tau_2 .
\end{array} \right.  
\end{equation}  
The value of the maximal efficiency $\etam$ is not well defined because of the lack 
of young pulsars in the Milky Way. The data seem to indicate that it is at least 0.3 \citep{kargaltsev08,vink11}, 
which we take as a lower limit. 
In principle, it can even exceed unity, because of the beaming of the pulsar radiation. 

In the $B$--$p$ plane, we can identify the lines of constant luminosity (see  solid lines in Fig.~\ref{f:lum_tau}). 
Depending on the range of $\tau_{\rm c}$, these lines have different slopes (see equation~(\ref{eq:luminos})):  
\begin{equation}\label{eq:b12_p}
B_{12}= \left\{  \begin{array}{ll}
5\times10^{-3}\ p^2_{-3}  L_{39}^{1/2} \eta^{-1/2} , & 
\mbox{if}  \  \tau_{\rm c} \leqslant \tau_1\  \mbox{or}\  \tau_{\rm c} \geqslant \tau_2 , \\
0.08 \ p^{4/3}_{-3}  L_{39}^{1/6} , & \mbox{if}\ \tau_1 \leqslant  \tau_{\rm c} \leqslant \tau_2 . 
\end{array} \right.  
\end{equation}  
 The line $L={\rm const}$ intersects  with the line $\tau_{\rm c}=\tau_1$ at a point 1 with 
 coordinates  $(p_{-3,1},B_{12,1})=(60 L_{39}^{-1/2} \etam^{3/4}, 18  L_{39}^{-1/2} \etam)$, 
 while an intersection with the line $\tau_{\rm c}=\tau_2$ occurs at point 2 
 $(p_{-3,2},B_{12,2})=(60 L_{33}^{-1/2}, 1.8  L_{33}^{-1/2})$.

As the pulsar period increases, its luminosity drops.   
If $B_{12}>B_{12,1}$ or $B_{12}<B_{12,2}$, a pulsar crosses the  line  of a given luminosity  
being at the constant efficiency branch $\eta=\etam$ or $\eta=10^{-4}$, while 
for $B_{12,2}<B_{12}<B_{12,1}$ it occurs at the decaying branch of $\eta$.  
Thus for the fixed magnetic field, the pulsar period at a given luminosity $L$ is
\begin{equation}\label{eq:period}
p_{-3} = \left\{  \begin{array}{ll}
14\ B_{12}^{1/2} L_{39}^{-1/4} \etam^{1/4} , & \mbox{if}\ B_{12}>B_{12,1} , \\
 6.7\ B_{12}^{3/4} L_{39}^{-1/8}, & \mbox{if}\ B_{12,2}<B_{12}<B_{12,1} , \\
 45\ B_{12}^{1/2} L_{33}^{-1/4}, & \mbox{if} \  B_{12}<B_{12,2}.
\end{array} \right.  
\end{equation}

\subsection{Steady-state distributions and the differential luminosity function}

\subsubsection{Steady-state period distribution}

The evolution of the distribution function of pulsars over the period $N(p)=dN/dp$ 
(for a given  magnetic field $B$) can be described by the following  evolution equation: 
 \begin{equation}\label{eq:evolution}
\frac{\partial N(p)}{\partial t} = - \frac{\partial}{\partial p} \left[ \dot p N(p) \right] + Q(p), 
\end{equation}  
with the source function describing the production of new pulsars per unit period and time given by 
 \begin{equation}\label{eq:source}
Q(p) = \dot N\ \frac{1}{p\ln 10}   G(\log p; \log \meanp, \sigmap) ,
\end{equation}  
and $\dot N = \int Q(p) dp$ is the total production rate per unit time. 
Equation (\ref{eq:evolution}) can be solved analytically and in the steady-state the solution 
takes the form: 
 \begin{equation}\label{eq:solution}
N(p)  =  \frac{1}{\dot p} \int_{0}^p Q(p') dp' .
\end{equation}   
It reduces to  
\begin{equation}\label{eq:n_p_erf}
N(p)  = \dot N \frac{p}{\alpha B^2}  
\left[ 1 +  \mathrm{erf} \left(\frac{\log (p/\meanp) }{\sigmap\sqrt{2}} \right) \right] 
\end{equation}  
for the source function given by equation (\ref{eq:source}). 
At periods which are much larger than the initial periods we get 
 \begin{equation}\label{eq:small_p}
N(p)  =   \frac{\dot N}{\dot p}  = 
3.2\times 10^{7}  B_{12}^{-2}\  \dot N_{\rm yr}\ p , 
\end{equation}   
where $\dot N_{\rm yr}$ is the pulsar birthrate per year.

If the magnetic field and the birth period distributions of the pulsars are lognormal, the steady-state 
period distribution averaged over the magnetic field distribution is given by 
\begin{equation}
\langle N(p) \rangle_B = \dot N  \ \frac{ e^{2\ln^2\!{10}\ \sigmab^2}}{\alpha \meanb^2} p 
\left[ 1 +  \mathrm{erf} \left(\frac{\log (p/\meanp) }{\sigmap\sqrt{2}} \right) \right] , 
\end{equation}
For the periods $p$ much larger than  $\meanp$, the distribution has a form:
\begin{equation}
\langle N(p) \rangle_B = 3.2\times 10^7\ \frac{e^{2\sigmab^2\ln^2\!{10}} }{\langle B_{12} \rangle^2} \, \dot N_{\rm yr}\  p ,
\end{equation}
where we used the relation 
$\langle B^\xi \rangle  
= \meanb^\xi \exp[(\xi\ \sigmab\ln 10)^2/2]$.

\begin{figure}
\centering
\includegraphics[width=0.95\linewidth]{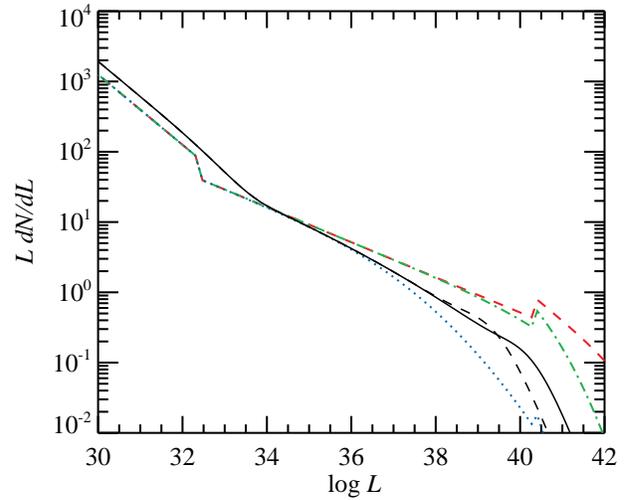}
\caption{Differential XLF of rotation-powered pulsars for the pulsar birth rate $\dot N_{\rm yr}=0.01$. 
Dashed red line shows the XLF for the fixed magnetic field $B_{12}=4$ and 
very small birth periods with $\log \meanp = -2.5$ and $\sigmap = 0.2$. 
The maximum efficiency is assumed to be $\etam=1$.
The dash-dotted green line corresponds to the larger birth periods with $\log\meanp = -2$, and 
the XLF for even larger $\log\meanp = -1.5$ is shown by the dotted line. 
The  XLFs averaged over the magnetic field distribution with parameters $\log\meanb = 12.6$ and $\sigmab = 0.4$ 
for the birth period distribution  parameters $\log\meanp = -1.7$ and $\sigmap = 0.2$  
are presented by the black solid and dashed lines for the maximum efficiency of $\etam=1$ and 0.3, respectively. 
}
\label{f:xlf}
\end{figure}

\subsubsection{Luminosity distribution}

For a given magnetic field, the XLF can be obtained from the period distribution function via transformation 
 \begin{equation}\label{eq:Ltransform}
 L\,N(L)  =  p\,N(p)\ \left| \frac{d\log p}{d\log L}\right| .
 \end{equation}   
For the power-law dependence of luminosity on period $L=C\ p^{-\gamma}$, we get 
 \begin{equation}\label{eq:XLF_p}
L N(L)  =   
3.2\times 10^7\ B_{12}^{-2}\  \dot N_{\rm yr}\  \frac{1}{\gamma} (L/C)^{-2/\gamma} . 
\end{equation}  
For the constant  X-ray efficiency $\eta$ (i.e. $\gamma=4$), 
we then easily get from equation (\ref{eq:luminos}) 
\begin{equation}\label{eq:XLF_const_eta}
L\, N(L)  =  5 \times  10^{3}\   B_{12}^{-1}\eta^{1/2}  \dot N_{\rm yr}\  L_{38}^{-1/2} . 
\end{equation}  
Thus for very young, rapidly rotating, luminous pulsars as well as for the old 
pulsars the distribution will follow that law (see Fig.~\ref{f:xlf}) 
In the intermediate regime for $ \tau_1<\tau_{\rm c}< \tau_2$, 
$\gamma=8$ and the XLF follows a shallower dependence:
\begin{equation}\label{eq:XLF_intermed}
L\, N(L)  =  3.2 \times  10^{2} \  B_{12}^{-1/2} \dot N_{\rm yr}\  L_{38}^{-1/4} . 
\end{equation}  
According to equation  (\ref{eq:luminos}), the high-luminosity break is 
expected at $L=4\times10^{41}B_{12}^{-2}\ergs$. 
Smaller mean periods and larger magnetic fields lead to a larger initial luminosity and 
therefore to a larger number of luminous sources. 
A corresponding low-luminosity break is at $4\times10^{33}B_{12}^{-2}\ergs$.

The examples of the XLF normalized to the pulsar birth rate $\dot N_{\rm yr}$ are presented in Fig.~\ref{f:xlf}. 
The XLF has a complex shape reflecting the behaviour of the X-ray radiative efficiency. 
The XLF for the fixed $B$ has sharp features reflecting breaks in the derivative 
of the function $L(p)$ given by equation (\ref{eq:luminos}). 
These breaks are unlikely to be observed, because the actual efficiency--age dependence (\ref{eq:xrayeff}) is likely to be smooth. 

The luminosity function of pulsars with the magnetic field distribution $G(\log B; \log \meanb, \sigmab)$ can 
be obtained by averaging the XLFs over that distribution.  In this case, the sharp features also disappear 
(see solid line in Fig.~\ref{f:xlf}). In the range of luminosities corresponding to the constant 
efficiency,  we get 
\begin{equation}\label{eq:XLF_eta_B}
L\, N(L)  =  5 \times  10^{3} \  \frac{e^{( \sigmab \ln\!{10})^2/2} }{\langle B_{12} \rangle}  
\eta^{1/2}  \dot N_{\rm yr}\  L_{38}^{-1/2} . 
\end{equation}  
In the intermediate range of luminosities, the power-law segment has the following form: 
\begin{equation}\label{eq:XLF_interm_B}
L\, N(L)  =  3.2 \times  10^{2}  \ 
\frac{e^{( \sigmab \ln\!{10})^2/8} }{\langle B_{12} \rangle^{1/2}}  
\dot N_{\rm yr}\  L_{38}^{-1/4} . 
\end{equation}  
If the birth period distribution has a peak at rather large periods, 
the XLF  has a cutoff before the high-luminosity power-law segment actually starts
 (e.g. see black solid line in Fig.~\ref{f:xlf}). 
Decreasing the maximum efficiency $\etam$ leads to a smaller cutoff  luminosity 
(compare solid and dashed black curves in Fig.~\ref{f:xlf}), while the 
intermediate power-law barely changes. 

Radiation from a pulsar may be confined within a narrow beam  $\sim 1$~str 
\citep{TM98} corresponding to the beaming factor $b=\Omega/4\pi=0.1$. 
However, young pulsars have PWN, which are more isotropic. 
The ratio of the observed luminosities of the nebula to the pulsar has a large 
spread, but typically is of the order unity \citep{kargaltsev08}. 
This  argues against strong beaming and therefore we take $b>0.3$.   
The normalization of the observed luminosity function 
scales  linearly with the beaming factor. 

\subsection{Non-stationary solution}

\subsubsection{Evolution of the period distribution}    

In order to determine the distribution of pulsars over the period at a given pulsar age,
we need to solve the time-dependent evolution equation (\ref{eq:evolution}):
\begin{equation}\label{eq:nonst}
\frac{\partial N(p,t)}{\partial t} + \frac{\partial}{\partial p}\big[\dot p N(p,t)\big] = 0
\end{equation}
with the following initial condition at zero-age:
\begin{equation}
 N(p, t=0)= N_0 (p).
\end{equation} 
According to equation (\ref{eq:p_t}), the period derivative can be expressed as $\dot p = \alpha B^2 /2p$. 
Equation (\ref{eq:nonst})  conserves the total number of pulsars $N$.  
Its solution is 
\begin{equation} \label{eq:solution2}
 N(p,t) = p \frac{N_0 (p_0) }{p_0}, 
\end{equation}
where $p_0 = \sqrt{p^2 - \alpha t B^2}$. The solution is only defined for 
\begin{equation}
 p \geqslant B \sqrt{\alpha t}.
\end{equation}
Function $N(p,t)$  has the mean 
\begin{equation}
 \langle p \rangle (t) = \int\limits_0^\infty \ p\ N(p, t)\ d p = 
 \int\limits_0^\infty \sqrt{p_0^2 + \alpha t B^2 }\ N_0 (p_0) dp_0,
\end{equation}
and the variance
\beq
 \sigma^2_p (t) & = &  \int\limits_0^\infty \left(p -  \langle p \rangle  \right)^2 N(p,t)\ dp  \nonumber\\
 & = & \int\limits_0^\infty \left( \sqrt{p_0^2 + \alpha t B^2} -  \langle p \rangle  \right)^2 N_0(p_0)\ dp_0 . 
\eeq
At large ages, the mean becomes 
\begin{equation}
\langle p \rangle  (t)
\approx \sqrt{\alpha t} B + \frac{\langle p_0^2 \rangle}{2 \sqrt{\alpha t} B } , 
\end{equation}
and the variance is  
\begin{equation}
\sigma^2_p (t)  \approx
\frac{\langle p_0^4 \rangle - \langle p_0^2 \rangle^2  }{4 \alpha t B^2}   \rightarrow 0.   
\end{equation}
Thus, the solution becomes the delta-function:
\begin{equation} \label{eq:phiv_delta}
 N(p, t \to \infty) = N\ \delta(p - \sqrt{\alpha t} B).
\end{equation}
This can be easily understood from equation (\ref{eq:p_t}), which describes the evolution of the pulsar period with time. 
At large age, the time-dependent term becomes much greater than the value of the initial spin period. 
Thus, every pulsar with a given magnetic field  at a given age has the same period $p(t) = \sqrt{\alpha t} B$. 
Therefore, at large ages the period distribution of the pulsars does not contain any information about the initial one. The characteristic timescale at which the information about the initial distribution is lost can be estimated as:
\begin{equation}
 t \approx \frac{1}{\alpha} \left( \frac{p_{0}}{B}\right)^2    
 \approx 1600 \left(\frac{p_{0,-2}}{B_{12}}\right)^2\,\,{\rm yr},
\end{equation}
where $p_{0,-2}$ is the birth period expressed in 10~ms.

\begin{figure}
\centering
\includegraphics[width=0.95\linewidth]{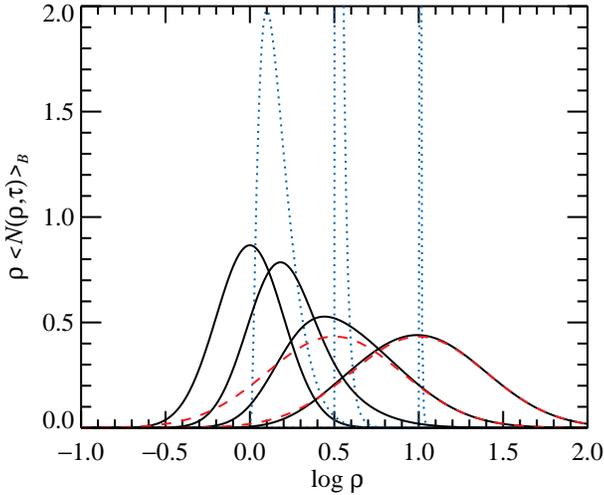}
\caption{Evolution of the (normalized) period distribution with time. 
The black solid lines show the evolution of period distribution with initial $\sigmap=0.2$
averaged over magnetic field distribution with $\sigmab=0.4$ 
for $\tau = 0$, 1, 10, 100 (from left to right). 
The blue dotted line shows the period distribution for a specific magnetic field ($\beta = 1$) at $\tau$=1, 10, and 100. 
The distribution becomes very narrow at large $\tau$, peaking at $\rho=\sqrt{\tau}$.  
The red dashed lines represent the asymptotic solution (\ref{eq:asympt}) at large $\tau$, 
which just reflects the lognormal distribution of the magnetic field. 
} 
\label{f:pfigure}
\end{figure}

We can now obtain the solution averaged over the magnetic field distribution
\begin{equation} \label{eq:phiv_B}
 \langle N(p, t)\rangle_B  =  \int\limits_{-\infty}^{\infty} N(p, t)\ G(\log B; \log \meanb, \sigmab)\ d\log B . 
\end{equation} 
It is useful to introduce dimensionless variables 
\begin{equation}
 \rho = \frac{p}{\meanp},\quad \beta = \frac{B}{\meanb}, 
 \quad \tau = \alpha \left(\frac{\meanb}{\meanp}\right)^2 t,
\end{equation}
and find the solution  
as a function of dimensionless period $\rho$ and time $\tau$, such 
as  $\langle N(\rho, \tau)\rangle_B=\langle N(p, t)\rangle_B \times \meanp$
and $\int \langle N(\rho, \tau)\rangle_B d\rho=N$. 
For the lognormal distribution of both magnetic field and birth periods,  we get 
\beq\label{eq:p_big}
\lefteqn{ \langle N(\rho, \tau)\rangle_B  
=  N \frac{\rho}{2\pi \sigmap \sigmab \ln^2 10 } }  \\
& \times & \!\!\! \int \limits_{0}^{\rho/\sqrt{\tau}} \!\!\!\! \frac{ d\beta}{\beta (\rho^2 - \tau \beta^2)} 
 \exp\left\{-\frac{\log^2\!\!\sqrt{\rho^2 - \tau \beta^2}}{2\sigmap^2} -  \frac{\log^2\beta}{2\sigmab^2}\right\} . \nonumber 
\eeq
An asymptote at large $\tau$ can be easy obtained directly substituting (\ref{eq:phiv_delta}) 
to equation (\ref{eq:phiv_B}): 
\begin{equation} \label{eq:asympt}
\langle N(\rho, \tau\to \infty)\rangle_B 
= \frac{N}{\rho \ln 10}   G(\log \rho; \log \sqrt{\tau}, \sigmab)
\end{equation}
This implies that the mean of the distribution $\rho N(\rho)$ increases with time as 
$\langle \rho \rangle = \sqrt{\tau}$  
and the dispersion is completely determined by the width of the magnetic field distribution $\sigmab$.
Evolution of the period distribution is presented in Fig.~\ref{f:pfigure}. 
If $\sigmap < \sigmab$, then at large $\tau$ the period distribution is wider than the initial one,
while in the opposite case, $\sigmap > \sigmab$, the period distribution becomes narrower. 

\begin{figure}
\centering
\includegraphics[width=0.95\linewidth]{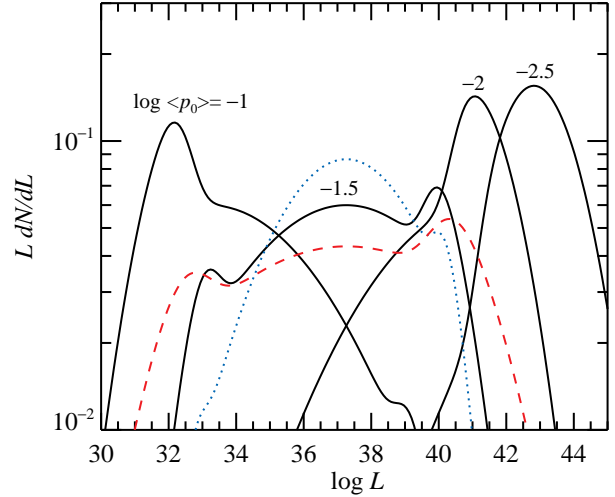}
\caption{Normalized luminosity distribution at birth for different values of birth period with $\sigmap = 0.2$, $\log\meanb = 12.6$,  $\sigmab = 0.4$ and $\etam=1$ (black lines). 
Blue dotted and red dashed lines show the luminosity distribution 
with $\sigmab = 0.2$ and  $\sigmap = 0.4$, respectively, for $\log\meanp = -1.5$ (other parameters as above). 
} 
\label{fig:birthlum}
\end{figure}

\begin{figure}
\centering
\includegraphics[width=0.95\linewidth]{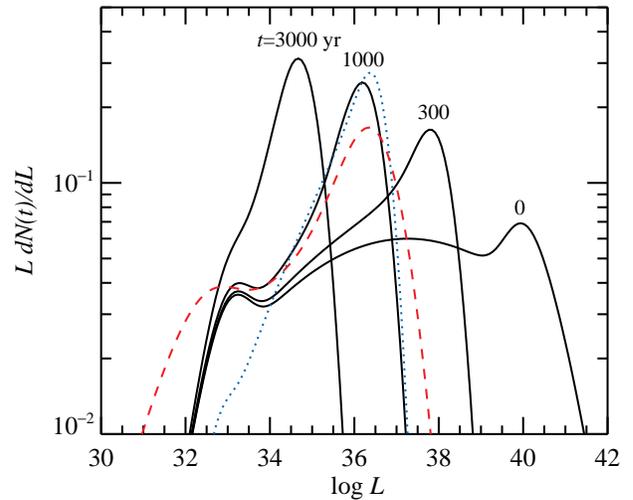}
\caption{Evolution of the  pulsar normalized luminosity distribution for the initial distribution with the following  
parameters $\log\meanp = -1.5$, $\sigmap = 0.2$, $\log\meanb = 12.6$,  $\sigmab = 0.4$ and $\etam=1$ (black solid lines).  
Blue dotted and red dashed lines shows the luminosity distribution at 1000~yr 
for $\sigmab = 0.2$ and $\sigmap = 0.4$, respectively (other parameters as above).  
The luminosity distribution becomes more symmetric at large age.
} 
\label{fig:lumevol}
\end{figure}

\subsubsection{Evolution of the luminosity distribution}
\label{sec:evol_lum}

In order to obtain the luminosity distribution, we take the time-dependent solution 
for the period distribution (\ref{eq:solution2}), use the  transformation (\ref{eq:Ltransform}) 
and average the derived expression over the magnetic field. 
Resulting distribution and its evolution is presented in Figs~\ref{fig:birthlum} and \ref{fig:lumevol}, respectively.  
As it is clearly seen from Fig.~\ref{fig:birthlum}, the initial luminosity distribution of the pulsars can be multimodal. 
Every mode of the distribution is related to the different regime of the luminosity-period dependence. 
Also, the luminosity distribution at birth may reveal the narrow spikes, related to the breaks in the $dp(L)/dL$ derivative.
The  luminosity distribution is broader for larger  $\sigmap$ and $\sigmab$. 
The distribution becomes  narrower and more symmetric as the time increases (Fig.~\ref{fig:lumevol}). 
This happens because at birth, pulsars can operate in the different regimes of  conversion of the rotational energy 
losses to the X-ray radiation, depending on the spin period and the magnetic field distributions. 
With time, all pulsars move towards the same regime, where the efficiency is constant $\sim10^{-4}$ (see equation (\ref{eq:xrayeff})).

\section{Observational constraints on model parameters}
\label{s:obser}

\subsection{Previous determination of magnetic field and birth period distributions}

Distributions of the pulsars over the magnetic field and the birth period were investigated in several papers based on 
the analysis of the observed radio \citep{arz02,kaspi06} and the gamma-ray pulsars \citep{gonthier02,takata11}. 
Parameters of the magnetic field distribution are similar in all these studies  lying in the range  $\log\meanb$=12.35--12.75, 
$\sigmab$=0.1--0.55 (see Table~\ref{t:par}). 
However, parameters of the birth period distributions are significantly different: 
the mean logarithm $\log\meanp$ varies from $-2.3$ to $-0.5$ 
(i.e. periods in the range from 5 to 200 ms) and the width $\sigmap$ varies in the range  0--0.8 (see Table~\ref{t:par}; 
the parameters were estimated by fitting the lognormal distribution to the actual distributions adopted by the authors). 
This difference in the birth period distributions  is most likely caused by a low sensitivity of the considered models to the birth period. As it was shown on Section~\ref{s:model}, the period distribution of the pulsars  at large time does not contain information about the birth periods. 
Therefore, in order to determine these parameters we have to use only young pulsars. 

 \begin{table}
\caption{Birthrates  and parameters of the magnetic field and the birth period distributions for 
rotation-powered pulsars. }
\label{t:par}
 \begin{tabular}{@{}llllllc}
  \hline
\# &   $\log\meanp$ & $\sigmap$ & $\log\meanb$ & $\sigmab$ & $\dot N_{\rm yr}$$^{a}$ &
 Reference$^b$ \\
\hline
1 & $-$2.3   & 0.3$^{c}$ & 12.35 & 0.4   & 0.0013 & 1  \\ 
2 & $-$1.52 & 0.0 & 12.75$^{d}$ & 0.33$^{d}$ & 0.01$^{e}$    & 2 \\ 
3 & $-$0.52$^{f}$ & 0.8$^{f}$ & 12.65 & 0.55 & 0.028$^{g}$ & 3 \\
4 & $-$1.7$^{h}$    & 0.1$^{h}$  & 12.6   & 0.1   & 0.01   & 4 \\
\hline
\end{tabular}
\flushleft
$^{a}$ Birth rate of pulsars in the Milky Way per year. \\
$^{b}$ References: (1) \citet{arz02};  (2) \citet{gonthier02}; (3) \citet{kaspi06}; (4) \citet{takata11}. \\
$^{c}$ Ref. 1 gives the lower limit on $\sigmap$ of 0.2. 
Taking a broader distribution with $\sigmap>0.3$ does not affect the results. \\
$^{d}$ Parameters for the lognormal distribution were estimated by fitting a more complex 
distribution adopted in ref. 2, see their Table 1 and eq. (1). \\
$^{e}$ Value from the first line of Table 8 of  ref. 2. \\
$^{f}$  Parameters for the lognormal distribution were estimated by fitting a Gaussian  
distribution adopted in ref. 3, see their Table 8. \\
$^{g}$ Birthrate from Table 8 of ref. 3. \\
$^{h}$ The lognormal distribution approximates the flat distribution in the 20--30 ms range adopted in ref. 4. \\
 \end{table}

Recently, \citet{popov12} have presented new estimates of the birth periods based on a sample of radio 
pulsars associated with the SN remnants.  They showed that the distribution has to be rather wide, 
and it is consistent with a Gaussian with the mean $p_0$$\sim$0.1~s and width $\sigma$$\sim$0.1~s. 
However, this result is inconclusive, because the number of objects in the used sample is not large enough to 
derive the exact shape of the period distribution.         

The analysis of the observed luminosity distribution of the historical core-collapse SNe by \citet{perna08} showed that  
the predicted number of bright pulsars in the  \citet{perna04} model is much larger than the observed number of luminous SNe. 
This discrepancy  is related to the assumed very short (5 ms) mean birth period  from \citet{arz02}. 
On the other hand, using parameters of the pulsar magnetic field distribution from \citet{kaspi06}, 
\citet{perna08} found that in order to satisfy the observed luminosity distribution of the historical core-collapse SNe, the birth period of the pulsars should be larger than 40--50~ms. 

In the following sections  we repeat the analysis by \citet{perna08} using a different efficiency-age dependence 
given by equation (\ref{eq:xrayeff}) as well as using the new data that became available after 2008. 
We also obtain constraints on the pulsar birth period distribution by comparing the simulated pulsar XLF  with  
the observed XLF of the bright sources in the nearby galaxies derived by \citet{mineo12}.

\subsection{Constraints from the luminosity distribution of core-collapse SNe}

\citet{perna08} proposed that constraints on the birth period distribution can be obtained by 
comparing the observed luminosity distribution of historical core-collapse SNe
with the simulated pulsar XLF (for the given ages), 
considering that the most probable remnant of the core-collapse SN explosion is a neutron star.    
  
\begin{table}
\caption{X-ray luminosities of historical SNe}
\label{t:lum}
\begin{tabular}{lrlc}
\hline
SN &  Age (yr)$^{a}$ & $\log L$ & References$^b$\\
\hline
1979C    & 26.8 & $38.43^{+0.06}_{-0.07}$ & 1 \\
1986E    & 19.6 & $38.15^{+0.13}_{-0.19}$ & 1 \\
1986J    & 21.2 & $38.93^{+0.02}_{-0.03}$ & 1 \\
1988Z    & 15.5 & $39.46^{+0.07}_{-0.08}$ & 1 \\
1990U    & 10.9 & $39.04^{+0.19}_{-0.26}$ & 1 \\
1994I    &  8.2 & $36.90^{+0.02}_{-0.04}$ & 1 \\
1995N    &  8.9 & $39.63^{+0.09}_{-0.11}$ & 1 \\
1996cr   &  4.2 & $39.28^{+0.08}_{-0.10}$ & 1 \\
1998S    &  3.6 & $39.58^{+0.05}_{-0.06}$ & 1 \\
1998bw   &  3.5 & $38.60^{+0.10}_{-0.11}$ & 1 \\
1999ec   &  5.9 & $39.49^{+0.05}_{-0.06}$ & 1 \\
2001em   &  4.7 & $40.76^{+0.08}_{-0.10}$ & 1 \\
2001gd   &  1.1 & $39.00^{+0.11}_{-0.15}$ & 1 \\
2001ig   &  0.5 & $37.54^{+0.20}_{-0.37}$ & 1 \\
2004C    &  3.1 & $38.00^{+0.11}_{-0.10}$ & 1 \\
2005ip   &  1.3 & $40.20^{+0.07}_{-0.09}$ & 2 \\
2005kd   &  1.2 & $41.41^{+0.06}_{-0.07}$ & 1 \\
2006jd   &  1.1 & $41.40^{+0.12}_{-0.17}$ & 3,4 \\
2008ij   &  0.56  & $39.00^{+0.11}_{-0.15}$ & 5 \\
\hline
\end{tabular}
\flushleft
$^a$ Ages of SNe were calculated from the detection times listed at 
the website of the IAU Central Bureau for Astronomical Telegrams, except for  SNe from \citet{perna08}. \\
$^b$ References: (1) \citet{perna08}; (2) \citet{immlerpooley07};  (3) \citet{immler07}; (4) \citet{dwarkadas2012}; 
(5) \citet{2008ij}. 
\end{table}

\begin{table} 
\caption{Upper limits for the X-ray luminosities of historical SNe  \citep[from ][]{perna08}.}
\label{t:uplim}
\begin{tabular}{lrl}
 \hline
SN & Age (yr) & $\log L$ \\
\hline
1923A    &  77.3 &  35.78    \\
1926A    &  75.3 &  37.15    \\
1937A    &  67.3 &  37.11    \\
1937F    &  62.1 &  36.43    \\
1940A    &  63.0 &  37.00    \\
1940B    &  62.6 &  36.93    \\
1941A    &  60.2 &  36.74    \\
1948B    &  55.1 &  35.67    \\
1954A    &  48.9 &  35.20    \\
1959D    &  41.6 &  37.34    \\
1961V    &  38.3 &  37.79    \\
1962L    &  41.2 &  37.67    \\
1962M    &  40.3 &  35.57    \\
1965H    &  37.7 &  38.18    \\
1965L    &  37.8 &  36.76    \\
1968L    &  32.0 &  36.18    \\
1969B    &  32.6 &  36.58    \\
1969L    &  30.3 &  37.68    \\
1970G    &  33.9 &  36.69    \\
1972Q    &  30.5 &  38.48    \\
1972R    &  31.9 &  35.86    \\
1973R    &  25.9 &  37.89    \\
1976B    &  26.2 &  37.95    \\
1980K    &  24.0 &  36.81    \\
1982F    &  22.6 &  36.04    \\
1983E    &  19.0 &  37.66    \\
1983I    &  17.8 &  36.23    \\
1983N    &  16.8 &  36.74    \\
1983V    &  19.1 &  37.85    \\
1985L    &  14.9 &  37.91    \\
1986I    &  17.1 &  38.48    \\
1986L    &  18.9 &  38.15    \\
1987B    &  14.1 &  38.18    \\
1988A    &  12.3 &  37.38    \\
1991N    &  11.8 &  37.62    \\
1993J    &   8.1 &  38.00    \\
1994ak   &   7.4 &  37.57    \\
1996ae   &   5.9 &  37.79    \\
1996bu   &   6.6 &  37.32    \\
1997X    &   6.1 &  37.34    \\
1997bs   &   2.5 &  38.46    \\
1998T    &   5.2 &  38.30    \\
1999dn   &   4.4 &  37.77    \\
1999el   &   5.6 &  38.75    \\
1999em   &   1.0 &  37.15    \\
2000P    &   7.2 &  39.08    \\
2000bg   &   1.3 &  39.15    \\
2001ci   &   2.5 &  37.70    \\
2001du   &   1.3 &  37.58    \\
2002ap   &   0.9 &  36.49    \\
2002fjn  &   4.7 &  39.11    \\
2002hf   &   3.1 &  38.88    \\
2003dh   &   0.7 &  40.70    \\
2005N    &   0.5 &  40.00    \\
2005at   &   1.7 &  38.48    \\
2005bf   &   0.6 &  39.78    \\
2005gl   &   1.6 &  39.53    \\
\hline 
\end{tabular}
\end{table}

One of the important questions is the earliest age at which SNe can be used to derive the observational luminosity distribution 
that would reflect the XLF of the brightest rotation-powered pulsars. 
There are two main issues here. The first problem is the high optical depth of the SNR shell at the earliest stages of its expansion. 
According to \citet{chevalier1994}, the optical depth of SNR changes with time as 
\begin{equation}
 \tau \sim \frac5{t_{\rm yr}^2},
\end{equation}
for typical parameters of the SN explosion (kinetic energy  $E_0\sim 10^{51}$~erg, mass of the ejecta 
$M_{\rm ej}\sim 10\,\msun$, and typical photon energy  $E \sim 10$~keV).
Therefore, the SNR shell becomes optically thin in about 3 years. 
However, \citet{perna08} limited their SNe subsamples by the minimal age of 10 and 30~yr. 
As a result, they did not include the most luminous SNe in their analysis. 
Furthermore, because of the finite size and rapid expansion of the shell, the diffusion time of photons in the SNR shell 
can be small enough to make the X-ray radiation of the central pulsar visible at even earlier times. 
Considering that the number of scatterings in the SNR shell of optical depth $\tau \gg 1$ scales as 
$N \sim \tau^2$, we can estimate the escape time of the photons from the shell: 
\begin{equation}
t_{\rm D}(t) \sim \frac{\tau(t)R_{\rm SNR}(t)}{c},
\end{equation}
where $R_{\rm SNR}(t)$ is the size of SNR.  
Because of the free expansion stage for SNe with ages $\lesssim 100$~yr, the size of SNR will increase with time as
\begin{equation}
 R_{\rm SNR}(t) = \sqrt{\frac{2 E_0}{M_{\rm ej}}}t.
\end{equation}
Therefore, the diffusion time is
\begin{equation}
 t_{\rm D} \approx \frac{0.05}{t_{\rm yr}}\,\,{\rm yr}.
\end{equation}
The diffusion time is equal to the age of SN at $t \sim 0.2$~yr and later it is always smaller.  
Therefore, a significant fraction of the SNR radiation may be produced by the central pulsar, 
because the luminosity of a typical newborn pulsar may achieve $10^{40}$--$10^{41}\,\ergs$, 
which is comparable to the highest observed SNe luminosities in our sample. 

The second issue is the fallback accretion onto a neutron star during early phases of SN explosion. 
According to \citet{chevalier89}, radiation from the central pulsar begins to diffuse through the accreting matter 
when a reverse shock radius reaches the radiation trapping radius. 
It happens at $\gtrsim 0.5$~yr after the SN.
Therefore, we can expect that the central pulsars will contribute to the total luminosity after 0.5--1.0~yr. 
This estimate is close to the limit coming from the diffusion time arguments. 
Thus, we will use the minimal age $t=0.5$~yr.

\begin{figure*}
\centering
\centering
\includegraphics[width=0.47\linewidth]{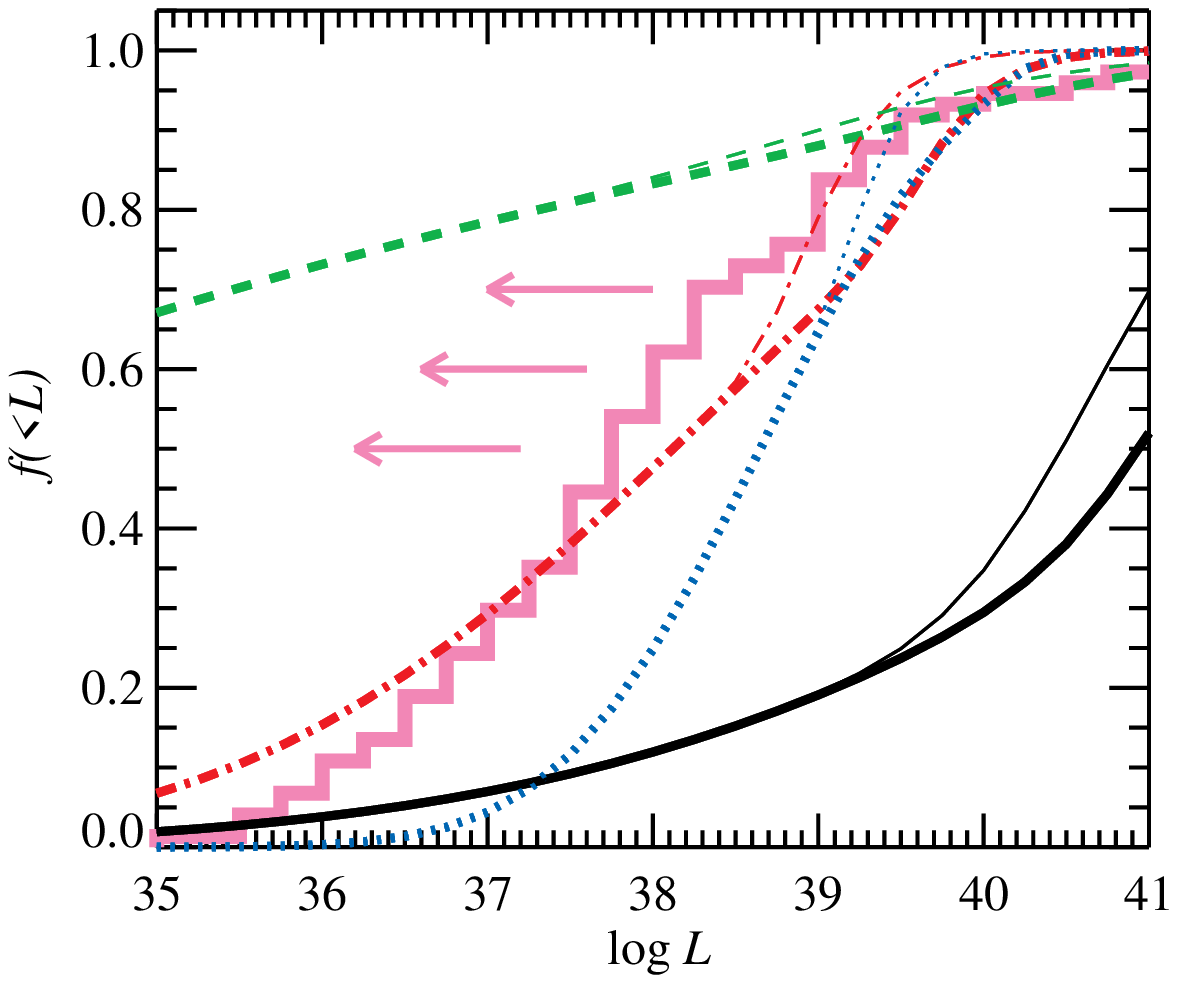}
\hspace{0.5cm}
\includegraphics[width=0.47\linewidth]{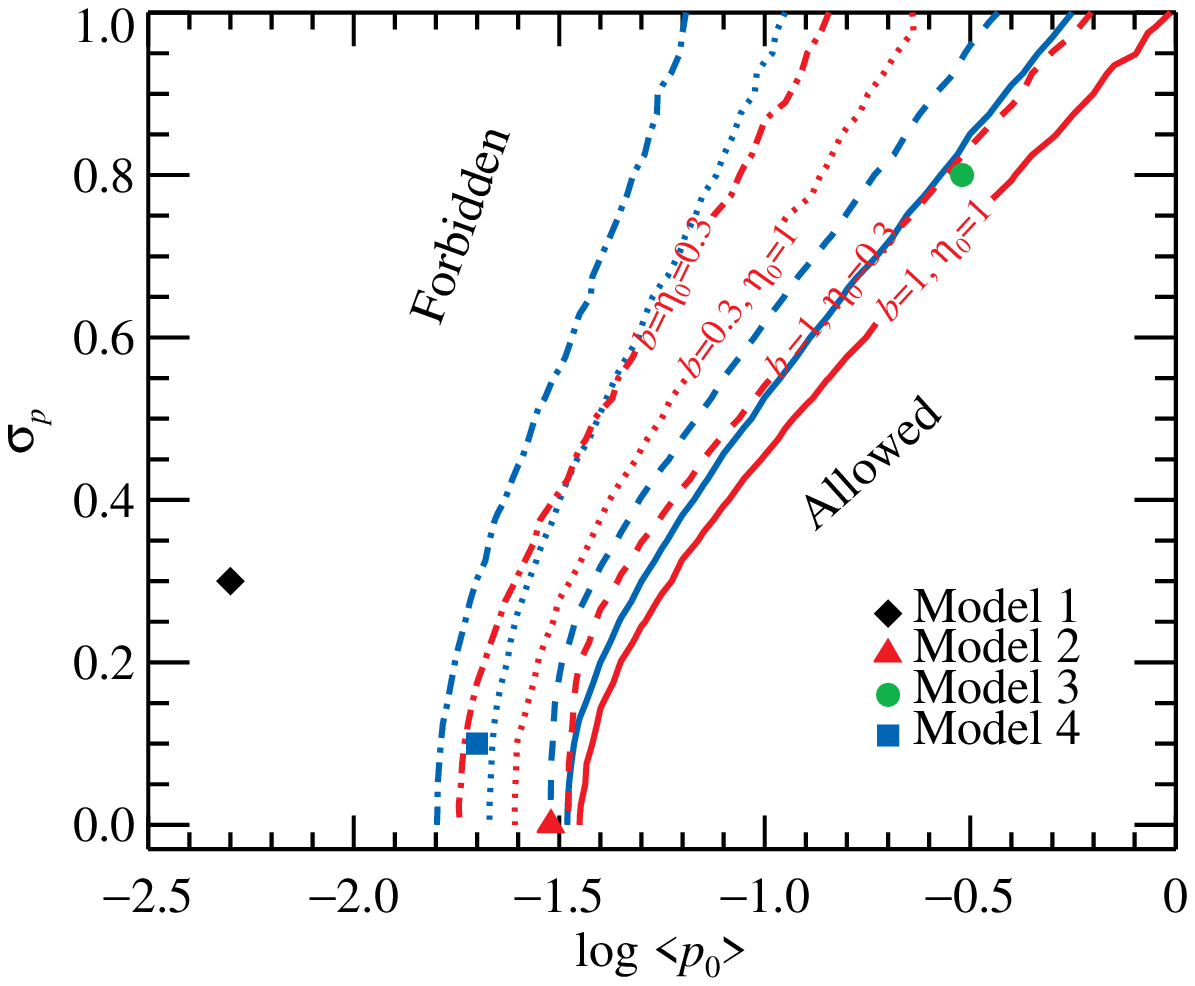}
\caption{(a) Cumulative normalized luminosity distributions for SNe  
(either measurements or upper limits; pink histogram)  with ages $t>0.5$~yr. 
Black solid, red   dash-dotted, green  dashed and blue dotted lines correspond to 
average distributions for models 1--4 from Table ~\ref{t:par}, respectively. 
Thick lines are for the case $\etam=1$ and the thin lines are for $\etam=0.3$.
Here no beaming is assumed ($b=1$).  
(b) Allowed region for the parameters of the birth period distribution. 
Parameters along the red and blue lines satisfy  condition (\ref{eq:supSNe}) 
in 90 and 68 per cent cases, respectively.  
The solid, dashed, dotted and dot-dashed lines correspond 
to different pairs of $(b,\etam)=$ (1,1), (1,0.3), (0.3,1) and (0.3,0.3), respectively.  
Regions to the right of these lines satisfy the data with higher probability. 
Calculations are performed for the average values  of the magnetic field distribution 
$\log\meanb = 12.6$ and $\sigmab = 0.4$. 
Positions of parameters listed in Table~\ref{t:par} are marked by different symbols. }
\label{f:snlumcomp}
\end{figure*}

Another important question is a selection effect, which may strongly affect the observed luminosity distribution of SNe, 
because the younger is the source the brighter it is and the higher is the probability for it to be detected. 
For example, most of the SNe luminosity measurements from \citet{perna08} are upper limits, because those sources are quite faint. 
Only 19 brightest and youngest sources have actual measurements of luminosity. 
Compilation of \citet{dwarkadas2012} contains additional eleven sources with known luminosity, 
but there are only four sources with ages $>0.5$~yr. 
In addition, because some fraction of the SNe X-ray luminosity is not related to the pulsar or PWN, 
also the actual X-ray  detections here should be treated as upper limits on the pulsar luminosity.  
For the analysis we use the data on the ages and the X-ray luminosities of core-collapse SNe (with ages $>0.5$~yr) 
from \citet{perna08} with the addition of the  new measurements from the compilation of \citet{dwarkadas2012} 
(see Tables~\ref{t:lum} and~\ref{t:uplim}).   
The cumulative histogram of upper limits is shown in Fig.~\ref{f:snlumcomp} by the bold pink line.

We follow the recipe described in Section~\ref{sec:evol_lum} 
and calculate the  luminosity distribution of 76 pulsars for the ages of SNe listed in Tables~\ref{t:lum} and \ref{t:uplim}. 
We then construct the average 
normalized cumulative distribution of luminosities $f(<L)$ and compare it to the observed  distribution. 
In the absence of beaming (i.e. $b=1$), 
from Fig.~\ref{f:snlumcomp} we see that only  model 3 \citep{kaspi06}  satisfies the upper limit distribution, 
for the maximum efficiency $\etam$ between 0.3 and 1.   
Model 2 \citep{gonthier02} is also reasonably close, especially for $\etam=0.3$. 

An additional effect appears if beaming is significant. 
Then most of the pulsars which appear to be faint in the X-rays, 
in reality could be very bright sources, but beamed away from us. 
Furthermore, about 10 per cent of SNe produce a black hole after explosion instead of a neutron star 
\citep{heger03}, which can be modelled as an additional multiplicative beaming factor 0.9.  
The cumulative normalized luminosity function in that case would 
start from the value $1-0.9b$ at the low-luminosity end.  
For example, for a smaller beaming factor $b = 0.3$, 
most of the bright pulsars would be undetected. 
In that case, models 2 and 3 satisfy the upper limit distributions, 
model 4 is only marginally consistent with them, but 
model 1 \citep{arz02}  still contradicts the data.  
 
We can also find more general constraints on the parameter set $(\log\meanp,\sigmap)$. 
With some high probability, the cumulative model distribution should 
be above the observed histogram of upper limits $f(<L)$ at any luminosity. 
We can formalize this condition by computing the fraction 
of model distributions that satisfy the constraint 
\begin{equation} \label{eq:supSNe}
\min_L \left\{ 
\frac{ f_{\rm pulsars}(<L) (\langle \log p_0 \rangle, \sigmap)  } 
{f_{\rm SNe}(<L)} \right\} > 1 . 
\end{equation}
Using Monte-Carlo method, we simulate 3000 sets of 76 pulsars at given ages 
(see Tables~\ref{t:lum} and \ref{t:uplim})
which follow given magnetic field and the initial period distributions, accounting 
for beaming and for the 10 per cent black hole fraction.  
We then find the dependence $\sigmap(\meanp)$, which satisfies  condition (\ref{eq:supSNe}) 
in 90 and 68 per cent cases. 
The results depend on  $b$  and $\etam$  (see  Fig.~\ref{f:snlumcomp}b). 
The mean birth period of the pulsars cannot be shorter than $\sim$15--30~ms for any $\sigmap$.
For larger dispersion $\sigmap \sim 1$, the limiting value is  between 60~ms and 1~s, 
depending on the parameters and the sought probability. 
The constraints depend much stronger on beaming than the assumed maximal efficiency $\etam$,
because they come mostly  from the low-luminosity SNe, whose number is not affected much by variation of $\etam$.
Our constraints on the mean periods are consistent with those derived by  \citet{perna08}, 
who  found $\meanp>$40--50~ms.

\begin{figure*}
\centering
\includegraphics[width=0.47\linewidth]{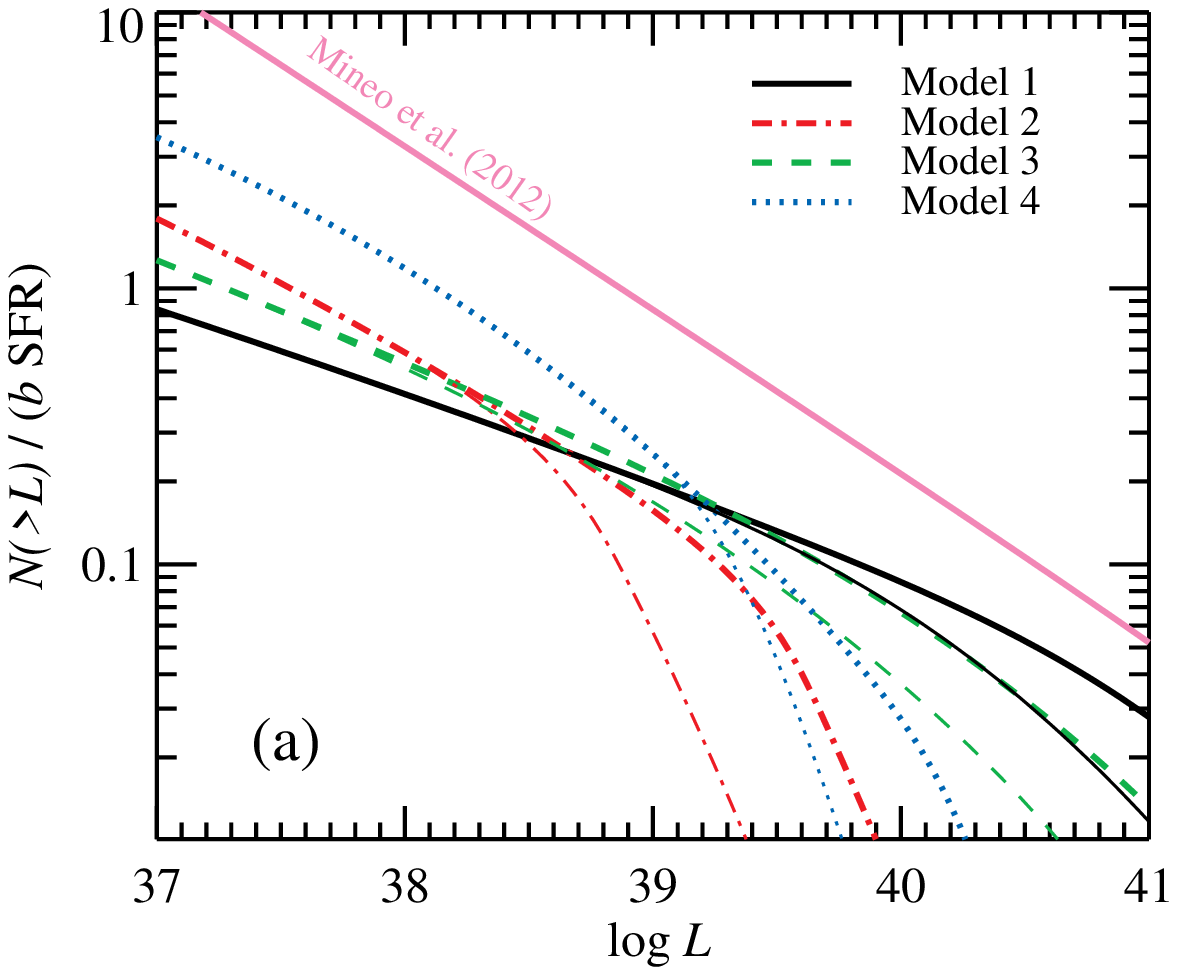}
\hspace{0.5cm}
\includegraphics[width=0.47\linewidth]{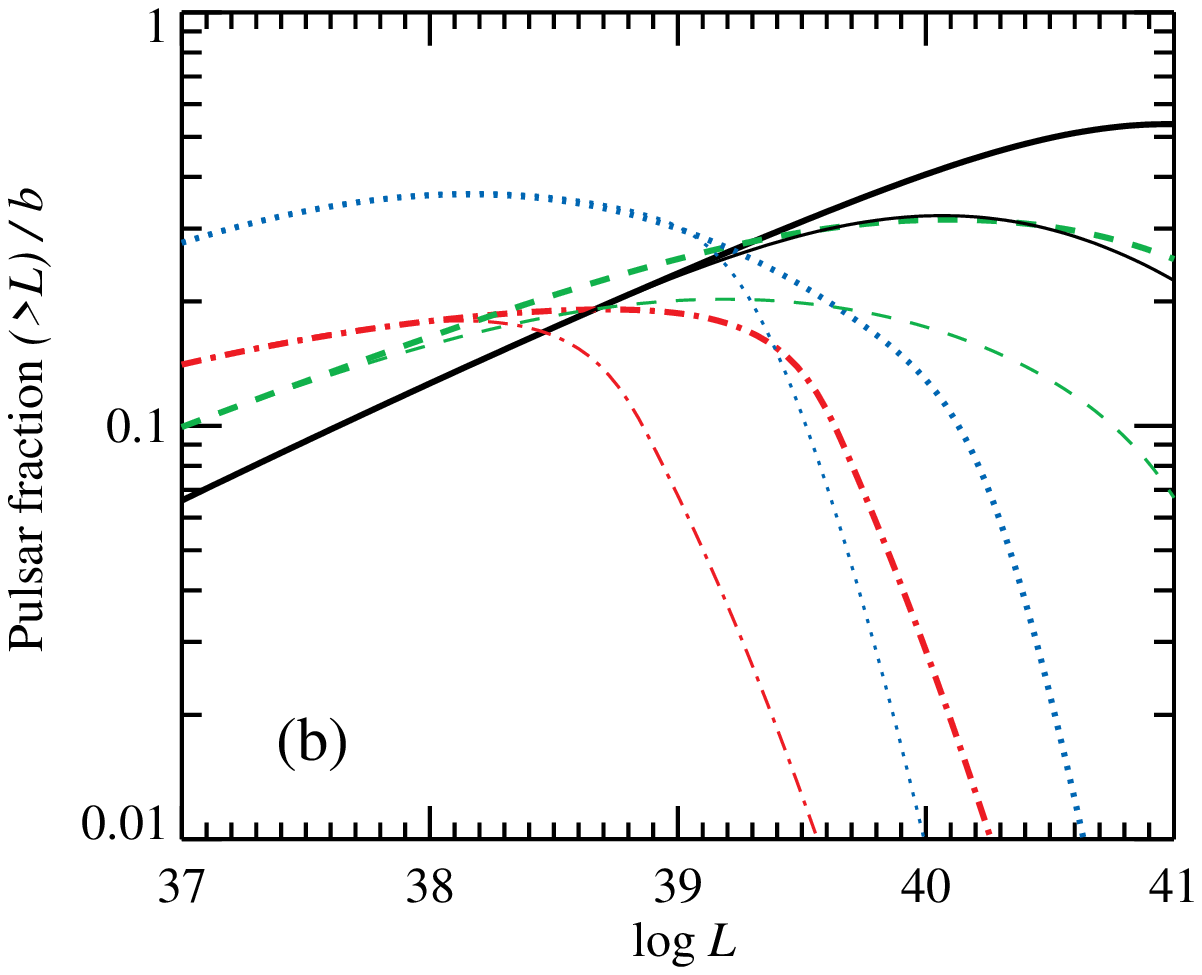}
\caption{(a) Cumulative luminosity distribution of the pulsars normalized to the 
unit SFR and unit beaming $b=1$. 
Black solid, red   dash-dotted, green  dashed and blue dotted lines correspond to models 1--4, respectively.  
The thick and thin lines correspond to the maximum efficiency  of $\etam=1$ and 0.3, respectively. 
The  pink solid straight line represents the average cumulative XLF 
of sources in the nearby galaxies \citep{mineo12} taken in the form (\ref{eq:mineo_cumul}).
(b) The predicted fraction (above a given luminosity) of pulsars in the high-luminosity end of the average XLF.   
Each line corresponds to the same models as in panel (a). 
}
\label{f:cumulative}
\end{figure*}

\subsection{Constraints from the  XLF for sources in nearby galaxies}\label{s:allowed}
\subsubsection{Averaged XLF of nearby galaxies}

The averaged XLF of the bright sources in nearby star forming 
galaxies was recently obtained by \citet{mineo12}.    
The star formation rates (SFRs) in the galaxies of their sample are spread in a broad interval between 
$\sim$0.1 and $\sim 100 \,\msun\,{\rm yr}^{-1}$. 
The observed XLF is well approximated by  a power-law:
\begin{equation}\label{eq:mineo}
 \frac{dN}{dL_{38}} = 1.88 \times L^{-1.59}_{38} \times \mbox{SFR} [\msun\,{\rm yr}^{-1}] . 
\end{equation}
\citet{mineo12} introduce the cutoff at $L_{\rm cut, 38} = 10^{3}$, 
because of lack of statistics at luminosities above $10^{41}\,\ergs$.  
Here we do not introduce the cutoff and integrate the
relation (\ref{eq:mineo}) to infinity to derive the cumulative distribution:
\begin{equation} \label{eq:mineo_cumul}
 N(>L_{38}) = 3.2 \times L_{38}^{-0.59} \times {\rm SFR}. 
\end{equation}

\begin{figure*}
\centering
\includegraphics[width=0.75\linewidth]{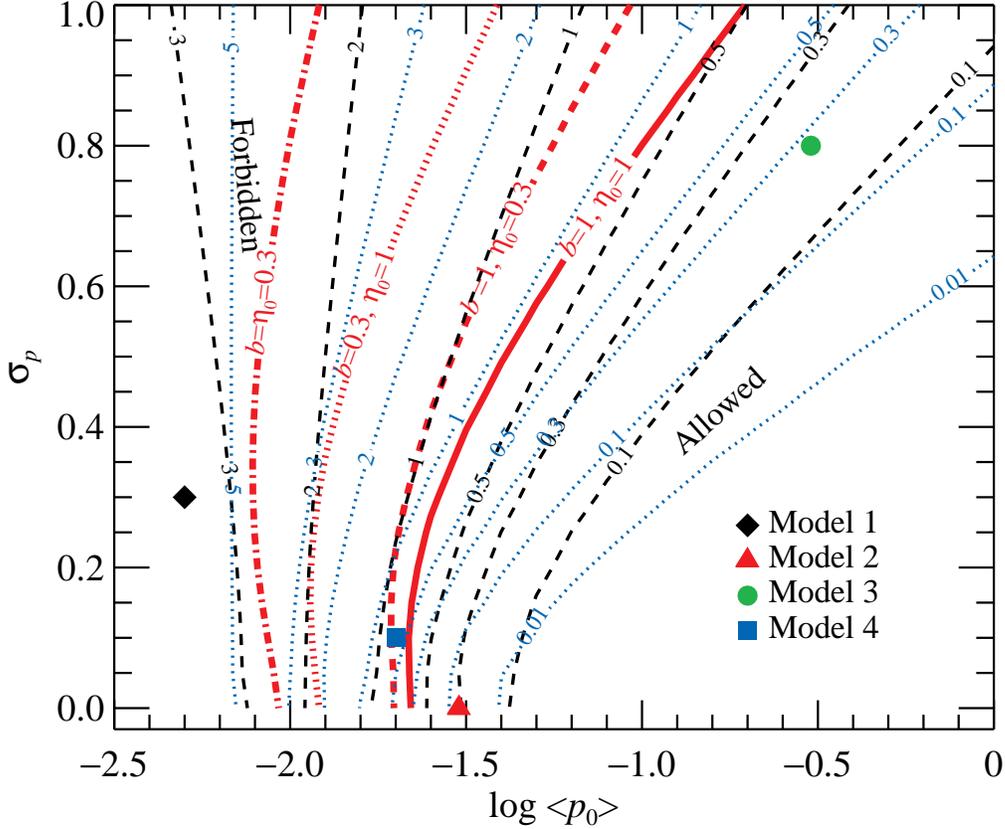} 
\caption{Allowed region for the parameters of the birth period distribution derived  
from the XLF of sources in the  nearby galaxies and the  
fraction of pulsars in the observed cumulative XLF \citep{mineo12}. 
Parameters along the bold red lines satisfy  equation (\ref{eq:sup}). 
The solid, dashed, dotted and dot-dashed lines correspond 
to different pairs of $(b,\etam)=$ (1,1), (1,0.3), (0.3,1) and (0.3,0.3), respectively.  
Regions to the right of these lines are allowed and to the left are forbidden. 
Dashed black and dotted blue contours give the pulsar fraction for $b=1$
at luminosities above $10^{39}\ergs$  and $10^{40}\ergs$, respectively. 
Calculations are performed for the average values  
$\log\meanb = 12.6$, $\sigmab = 0.4$  and $\kappa=0.01$. 
Positions of parameters listed in Table~\ref{t:par} are marked by different symbols. }
\label{f:allowed}
\end{figure*}

\subsubsection{Comparisons of the pulsar and observed XLF}
\label{s:XLF_diff}

In order to make the comparisons between the pulsar XLF  and the 
observed XLF of sources in the nearby galaxies, 
we first need  to find the relation between the pulsar birthrate and the SFR. 
We assume the Galactic SFR$_{\rm MW}=2\,\,\msun$ yr$^{-1}$,  in accordance with the recent study of \citet{chomiuk2011}. 
However, using luminous radio SN remnants and the X-ray point sources, these authors 
found that the Milky Way deviates from SFR expectations at the 1--3$\sigma$ level, 
hinting that  the Galactic SFR is overestimated or extragalactic SFRs need to be revised upward.

The estimations for the birth rate of pulsars in the Milky Way differs by an order of magnitude in various papers 
(see Table~\ref{t:par}), and on average is about 0.02 yr$^{-1}$.  
  The conversion between pulsar birthrate and the SFR can be expressed as follows : 
\begin{equation}\label{e:nSFR}
 \dot N \left[{\rm yr}^{-1} \right] = \kappa \times \mbox{SFR}  \left[ \msun \  {\rm yr}^{-1}\right],
\end{equation}
with the conversion factor $\kappa$ varying between 0.0007 (in model 1) 
and 0.014 (in model 3), with the mean of about 0.01.

Using the conversion formula (\ref{e:nSFR}) we can now produce the cumulative luminosity distribution of  pulsars 
normalized by the SFR and compare it to the observed XLF from \citet{mineo12}. 
The XLFs calculated for the four models from Table~\ref{t:par} are presented in Fig.~\ref{f:cumulative}(a). 
We see that all cumulative XLF are harder than the observed XLF at luminosities below $10^{38}\ergs$.
This fact is easy to understand from our Fig. \ref{f:xlf} and equation (\ref{eq:XLF_intermed}): 
the typical slope of 1.25 is related to the intermediate characteristic ages 
(see eq.[\ref{eq:xrayeff}]), where the efficiency varies strongly. 
The position of the cutoff depends not only on the mean birth period, but also on the width of the 
distribution. For example, model 3 has the largest mean period, 
but because of a large dispersion, the  XLF   extends to very high luminosities without a visible break.
On the other hand, model 2 has a rather small mean period, but the XLF 
cuts off sharply, because of the zero $\sigmap$ and the absence of fast pulsars. 
Model 1 has the largest number of bright pulsars because of the smallest mean period (see also Fig. \ref{f:xlf}). 
Model 4 also shows a cutoff at rather small luminosity in spite of the small initial periods, because 
of the narrow magnetic field distribution.

The number of high-luminosity pulsars depends on the maximum efficiency $\etam$. 
Decreasing $\etam$ leads to a smaller cutoff  luminosity 
(compare thick and thin curves in Fig.~\ref{f:cumulative}(a)). 
However, if the cutoff is at very large luminosity (as e.g. in the case of models 1 and 3), 
variations in   $\etam$ do not affect significantly the observed XLF, 
at least in the range of luminosities  $\log L<40$.  
The XLF normalization scales linearly with the beaming factor $b$. 

Dividing the pulsar cumulative XLF by the observed XLF, 
we obtain the fraction of pulsars as a function of luminosity. 
It is an increasing function of luminosity and reaches the maximum 
at $\log L$ between 38 and 41, depending on the model parameters. 
The maximum pulsar fraction reaches (0.2--0.5)$b$ for all models.

\subsubsection{Constraints on the birth period distribution}

The differential XLF of the pulsars should not exceed the observed XLF at any luminosity. 
This condition gives us the opportunity to find constraints on the birth period distribution.
We can find the dependence $\sigmap(\meanp)$, which satisfies the condition
\begin{equation} \label{eq:sup}
\max_{L < L_{\max}} \left\{ 
\frac{ \mbox{XLF}_{\rm pulsars} (L, \langle \log p_0 \rangle, \sigmap)  } 
{\mbox{XLF}_{\rm Mineo}(L)} \right\} = 1,
\end{equation}
where $L_{\max} = 10^{41}\ergs$ corresponds to the maximal  luminosity considered by \citet{mineo12}. 
As with the constraints from the SNe, here 
the results depend on the assumed values of $\etam$ and $b$ (see  Fig.~\ref{f:allowed}). 
As we  see, the mean birth period (for  $b=1$ and $\etam=1$) 
has to be larger than 25--250~ms, depending on the width of the period distribution. 
Parameters considered by \citet{arz02} and \citet{takata11} lie in the forbidden region. 
Parameters from other papers listed in Table~\ref{t:par} are in the allowed region. 
However, as it was  shown by \citet{popov12}, the period distribution has to be rather wide and to cover the range of periods 
from tenths to hundreds of milliseconds. Only the distribution found by \citet{kaspi06} satisfies this condition. 

For a smaller value of the  maximal efficiency $\etam= 0.3$, 
the critical line  (red bold dashed line in Fig.~\ref{f:allowed}) 
shifts to the left and depends weaker on $\sigmap$. 
Thus, the decrease of the efficiency leads to shorter allowed periods. 
Variations in the beaming factor lead to a stronger effect. 
For $b=0.3$ and $\etam= 1$, the allowed region extends beyond the parameters from \citet{takata11}, but still 
cannot reach the parameters from \citet{arz02}. 
 
We note here that  for simulations we used the average value of the pulsar birth 
rate $\kappa=0.01$, while it is more than 10 and 2 times smaller in  models of  \citet{arz02} and \citet{takata11}, 
respectively.  Thus all considered models are in principle allowed if one corrects for different $\kappa$.   
However, for small $b$ and $\etam$ the constraints coming from the SNe (see Fig.~\ref{f:snlumcomp}b) are 
actually stronger and rule out model 1, with model 4 being only marginally consistent with the data.

\begin{figure*}
\centering
\includegraphics[width=0.47\linewidth]{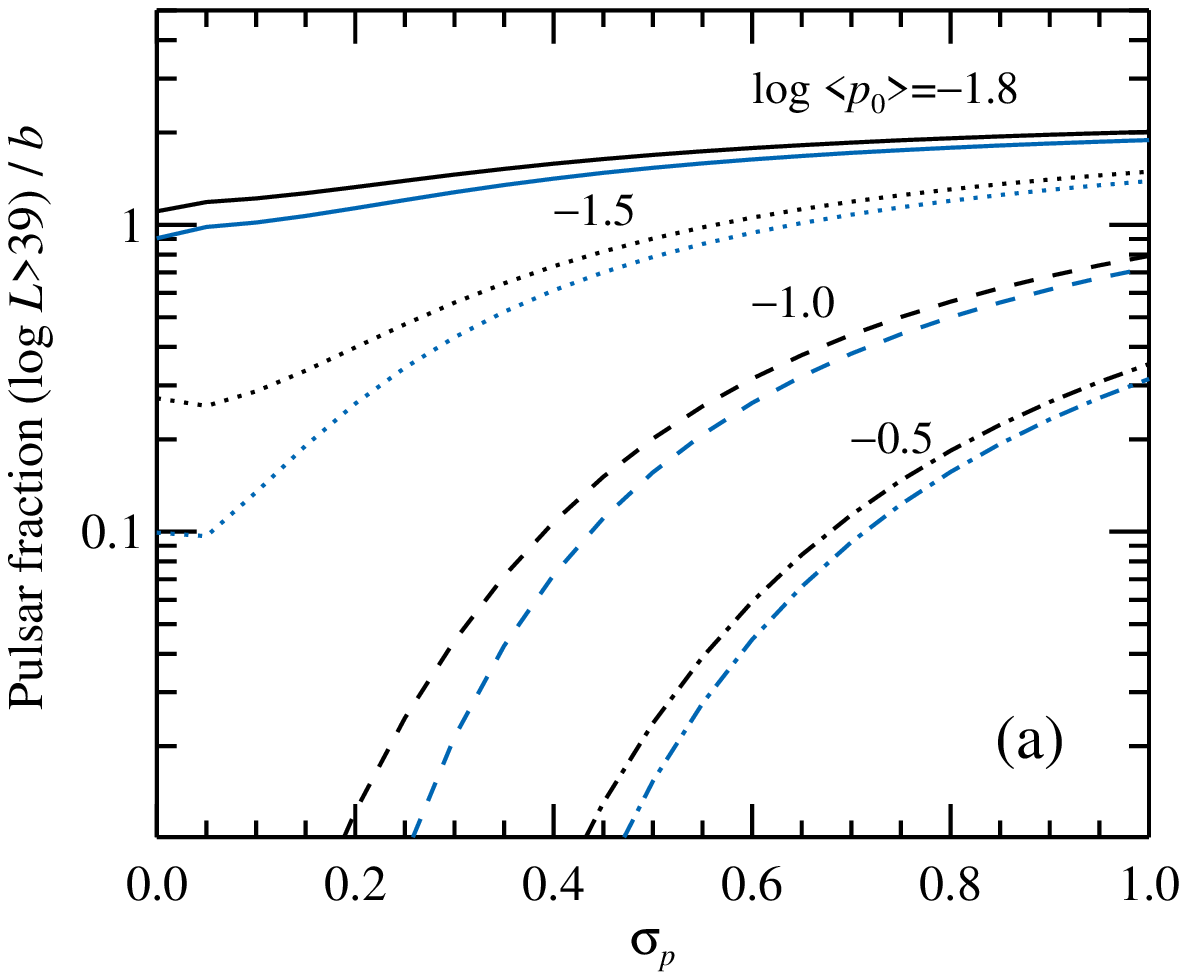} 
\vspace{0.5cm}
\includegraphics[width=0.47\linewidth]{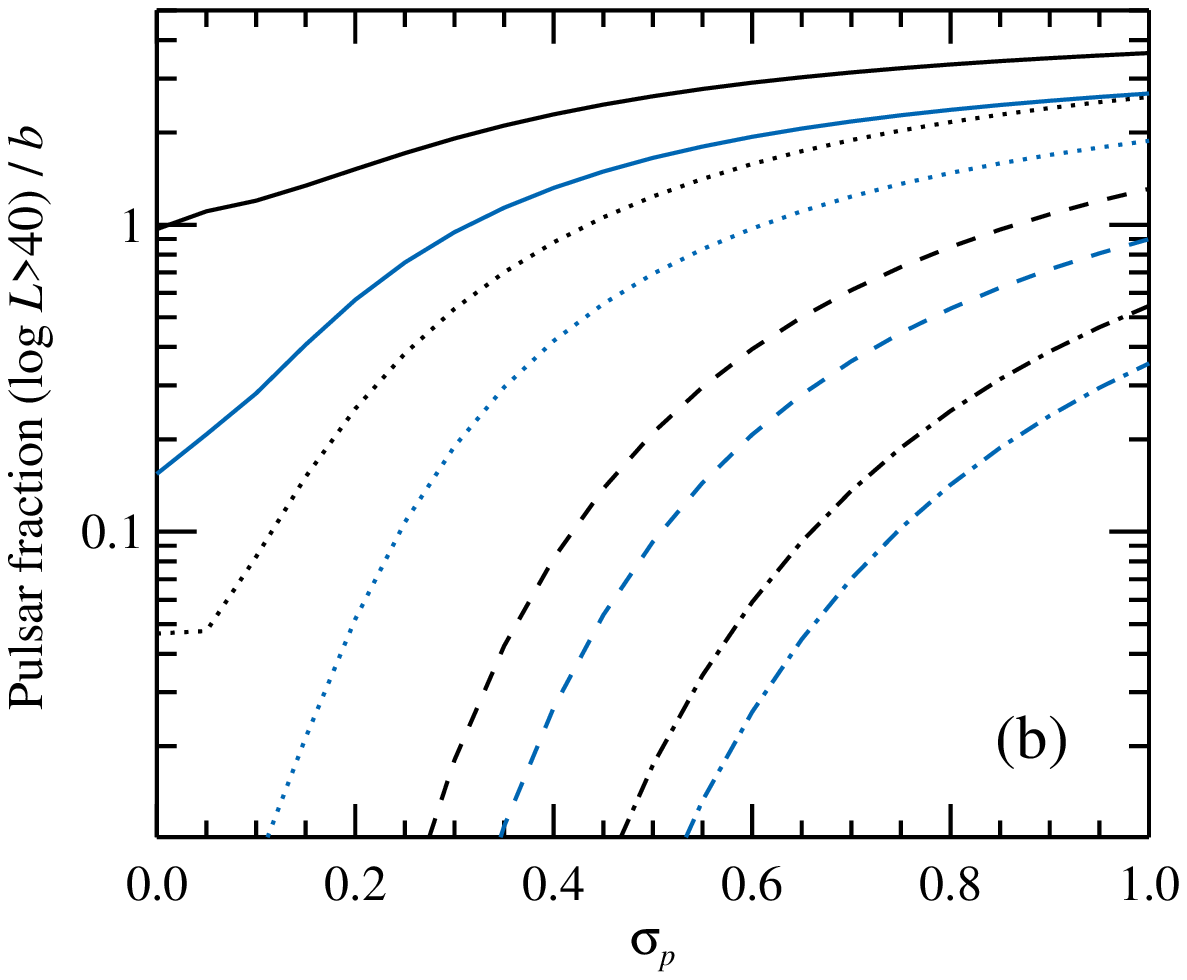} 
\caption{The fraction of the pulsars in the observed cumulative XLF \citep{mineo12}  
at luminosities above $10^{39}\ergs$ (panel a)  and above $10^{40}\ergs$ (panel b) (see Section~\ref{s:frac}) 
as a function of $\sigmap$.
Solid, dotted, dashed and dot-dashed lines correspond to the mean period of $\log\meanp=-1.8, -1.5, -1.0$ and 0.5, 
respectively. 
The upper black and the lower blue lines give the pulsar fraction for $\etam=1$ and $0.3$, respectively.  
Calculations are performed for the values  $\log\meanb = 12.6$, $\sigmab = 0.4$ and $\kappa=0.01$. 
}
\label{f:fraction}
\end{figure*}

\section{Pulsar contribution to ULX} 
\label{s:ULX}

\subsection{Dependence on the birth period distribution}
\label{s:frac}

The total number of the luminous pulsars with luminosities greater that $10^{39}\ergs$ is very similar 
in all four models 1--4 (see Fig.~\ref{f:cumulative}(a)): 
\begin{equation}
 N_{\rm obs}(\log L > 39) \approx 0.3\times b \times \mbox{SFR}\,\big[\msun \,\, {\rm yr}^{-1}\big]. 
\end{equation}
At  larger luminosities this number depends strongly on the period distribution and the maximum efficiency. 
For example, model 2 predicts less than $0.01b \times \mbox{SFR}$ pulsars 
above $10^{40}\ergs$ because of the cutoff at  $\sim2\times10^{39}\ergs$ in the XLF (see Fig.~\ref{f:cumulative}(a)). 
On the other hand, model 3 gives about $0.08b \times \mbox{SFR}$ very bright pulsars. 
The pulsar fraction  in the observed XLF at $\log L > 39$ can be as high as (0.2--0.3)$b$ for all models.
At even higher luminosities,  this fraction drops in models 2 and 4 and increases in models 1 and 4.

We can also calculate the pulsar fraction dependence on the combination $(\log \meanp,\sigmap)$. 
This fraction calculated for sources with luminosities $\log L > 39$ and 
$>40$ is presented as contours in Fig.~\ref{f:allowed}.
We see that the models 2 and 3 lie nearly on the same curve. 
The explicit dependence of the pulsar fraction on the parameters of the birth period distribution 
is shown in Fig.~\ref{f:fraction}.
We see that for small initial mean periods, the pulsar fraction is nearly independent 
of $\sigmap$ because the XLF cuts off at very high luminosities. The situation changes dramatically 
at large $\meanp$: the narrow period distribution now predicts cutoff at low luminosity and 
the pulsar fraction is negligible. At large widths $\sigmap\sim 1$, the pulsar fraction is still large 
because of the large extent of the XLF. 
Situation is similar for the cutoff luminosity $\log L=40$, but now for 
small $\sigmap$ the XLF cuts off  close to the limiting luminosity even for rather short initial periods 
and the pulsar fraction is small in that case. 
For large $\sigmap\sim1$, the pulsar fraction exceeds that fraction for  $\log L=39$  if $\log \meanp\lesssim-0.5$.

\subsection{Dependence on the maximum efficiency}

In a general case, the dependence of the pulsar fraction above   $\log L=39$ 
on $\etam$ can be easily seen in Fig.~\ref{f:fraction}(a). 
For small initial periods, the pulsar fraction is nearly independent of $\etam$ for all $\sigmap$ because 
the XLF extends to very high luminosities. 
For larger initial mean periods, the dependence on  $\etam$ is strong for narrow distributions $\sigmap\sim0$, 
because the XLF has a sharp cutoff around   $\log L=39$. For broad initial distributions with $\sigmap\sim1$, 
the pulsar fraction is still rather large and the dependence on $\etam$ is not so strong.  

The pulsar fraction at luminosities  in excess of $\log L=40$ (see Fig.~\ref{f:fraction}(b))  
shows a similar behaviour, but dependence on $\etam$ is stronger because 
typically the XLF cuts off at that luminosity even for small initial periods and large $\sigmap$.  

\begin{figure}
\centering
\includegraphics[width=0.9\linewidth]{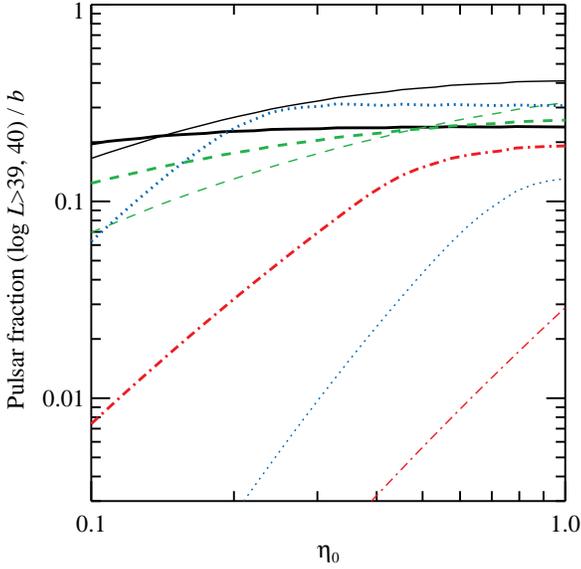} 
\caption{The predicted fraction of pulsars at luminosities above $10^{39}\ergs$ 
(thick lines)  and above $10^{40}\ergs$ (thin lines) 
in the average XLF of sources in the nearby galaxies \citep{mineo12} taken in the form (\ref{eq:mineo}) 
as a function of the maximum efficiency. 
Black solid, blue dotted, red dot-dashed  and green dashed curves correspond to 
models 1--4 from  Table~\ref{t:par}, respectively. 
}
\label{f:ULXeta}
\end{figure}

The dependence  of the pulsar fraction on the maximum  efficiency $\etam$ 
for the models listed in Table~\ref{t:par} is presented in Fig.~\ref{f:ULXeta}.
Most of the models predict rather flat dependence  on $\etam$ of the pulsar fraction at 
$\log L>39$ because of the wide initial period distribution producing the XLF extending to rather high luminosities.  
The only exception is model  2 \citep{gonthier02}, which predicts a significant drop in the pulsar function 
below $\etam=0.5$. This is a direct consequence of the fact that this model 
has a narrow period distribution and its XLF has a sharp cutoff at  about $\log L=39.3$ for $\etam=1.0$. 
Thus we see that for a rather wide range of $\etam$ between 0.3 and~1 
the pulsar fraction above $\log L=39$ is between about 10 and 30 per cent for all models. 
Obviously, the beaming can reduce this fraction proportionally and for $b=0.3$ it is then at least 3 per cent.

For models 1 and 3, the pulsar fraction is even larger at very high luminosities in excess of   $\log L=40$ 
reaching $0.4b$ and $0.3b$, respectively.  
In those cases, the dependence on $\etam$ is also not strong. While for model 2, 
the pulsar fraction is below 3 per cent and scales approximately as $\etam^2$.
From Fig.~\ref{f:allowed} it is clear that the closer parameters of the birth periods 
are to the limiting (bold red) line, the larger is the pulsar fraction. 
For large $\sigmap$,  the cumulative XLF is less steep than the observed XLF and therefore 
the pulsar fraction is a monotonically growing function that  can reach 100 per cent above  $\log L=40$. 
Thus it is possible that the pulsar fraction among the brightest ULX is significantly larger than 10 per cent.

\subsection{Distribution functions of the luminous pulsars}

\begin{figure}
\centering
\includegraphics[width=0.9\linewidth]{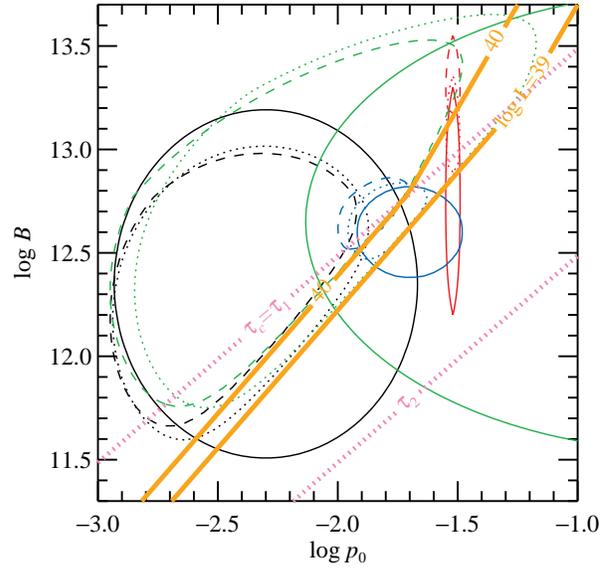} 
\caption{Distribution of pulsars over magnetic field and initial periods. 
Black,  red, green and blue solid curves encircle 90 per cent of pulsars 
for  models 1--4 from  Table~\ref{t:par}, respectively.
The dotted and dashed curves  encircle 90 per cent of initial pulsar distribution
that can be observed to radiate above $10^{39}\ergs$ and $10^{40}\ergs$, respectively. 
The mean and the standard deviation describing these distributions are given in Table~\ref{t:luminous}.  
Dotted pink curves give the dependence $\tau_{\rm c}=\tau_1$ and $\tau_{\rm c}=\tau_2$. 
Solid brown curves are the lines of constant X-ray luminosity of $10^{39}$ and $10^{40}\ergs$ 
for $\etam=1$.
}
\label{fig:distrib}
\end{figure}

In order to describe the possible observational appearance of the pulsars that can be observed as ULXs, 
we find a posteriori distribution of pulsars with luminosities $\log L> 39$ and $40$
over magnetic field and birth periods as well as over their true ages.
Because the pulsar luminosity drops with time, we are interested only in pulsars 
that emit above a given limiting $L$  at birth.
On the $\log B$--$\log p_0$ plane, these are the pulsars to the left of the corresponding 
$L={\rm const}$ line  (see Figs~\ref{f:lum_tau} and \ref{fig:distrib}). 
The probability that a pulsar will be observed above 
a given luminosity  threshold is proportional to the pulsar age when it crosses the limiting $L={\rm const}$ line. 
Thus the density distribution of such pulsars on the $\log B$--$\log p_0$ plane 
(limited to the region left of the limiting luminosity line) 
is given by the product of the  density  distribution at birth  and the true age  
$t=  (p_{\rm c}^2 - p_0^2)/\alpha B^2$: 
\begin{eqnarray}\label{eq:prob_lumin}
P(\log p_0, \log B) & \propto&  t \times  H(p_{\rm c} - p_0) \times G(\log p_0; \log \meanp, \sigmap)   \nonumber \\
& \times &   G(\log B; \log \meanb, \sigmab) ,
\end{eqnarray}
where $H$ is the Heaviside step function and 
$p_{\rm c}$ is the pulsar period 
when it crosses the limiting luminosity line given by equation (\ref{eq:period}).

These density distributions of the bright observed pulsars  for various models from Table~\ref{t:par} 
are shown in Fig.~\ref{fig:distrib}. 
These distributions are generally narrower than the original distribution and skewed towards 
smaller periods and larger magnetic fields. They are elongated along the line of constant 
luminosity.
The mean values and the standard deviations of these distributions are given in  Table~\ref{t:luminous}. 
We also find the  distribution of true ages of these pulsars, which is a monotonically decreasing function and can 
be described by the median age $\tau$.

The evolution of the luminosity for the average luminous pulsar can be described as follows:
during the first ten years the luminosity is nearly constant at the level of $\sim 10^{40}\,\ergs$. 
After that it starts to decrease and still exceeds $\sim 10^{39}\,\,\ergs$ for the next 100 years,
during which  the pulsars can be observed as ULXs. 
After that the luminosity decreases down to $\sim 10^{36}\,\,\ergs$ in about 1000 yr.

\begin{table}
\caption{Parameters of the  birth period and the magnetic field 
distributions as well as the median true age  of  pulsars with observed luminosities in
excess of a given value  for models from Table~\ref{t:par}. 
}
\label{t:luminous}
\begin{tabular}{cllllr}
 \hline
Model &  $\log\meanp$ & $\sigmap$ & $\log\meanb$ & $\sigmab$ &  $\tau$  \\
      &   & & & &  yr  \\
\hline
\multicolumn{6}{c}{$\log L>39$} \\
1 & $-$2.43 & 0.24 & 12.35 & 0.34   & 343 \\
2 & $-$1.52 & 0.0   & 13.13   & 0.16   & 96 \\
3 & $-$2.05 & 0.39 & 12.79   & 0.44   & 136 \\
4 & $-$1.80 & 0.08 & 12.67   & 0.09   & 105 \\
\multicolumn{6}{c}{$\log L>40$} \\
1 & $-$2.46 & 0.23 & 12.39  & 0.32  & 174 \\
2 & $-$1.52 & 0.0   & 13.37  & 0.13  & 13  \\
3 & $-$2.22 & 0.34 & 12.71  & 0.41  & 52 \\
4 & $-$1.86 & 0.07 & 12.70  & 0.08  & 36 \\
\hline
\end{tabular}
\flushleft
\end{table}

The rotation-powered pulsars are often assumed to be non-variable sources, but there may exist some variability on the 
time scales  shorter that $\sim 100$~yr related to the interaction of the SN remnant with the PWN and the surrounding media. 
As it was shown by \citet{dwarkadas2012}, the SN remnant could show a variability at least on the time scales $\sim 10$~yr. 
This variability could depend on the scale and the spatial spectrum of inhomogeneities of the surrounding media, 
and the characteristic variability timescale may increase with the pulsar age. 
The spectrum in the 0.1--10~keV range should consist of the soft thermal component related to the shock  
and the power-law tail related to the synchrotron radiation both from the shocks and the central pulsar.

\section{Summary}
\label{s:summary}

In the present paper we have investigated the question whether rotation-powered pulsars and PWN could be 
observed as some subclass of ULXs, and, if it is so, what is the fraction of pulsars in the whole ULX population.  

First, we developed an analytical model of the X-ray luminosity function, 
by solving the evolution equation for the period distribution of the pulsars. 
We derived both the steady state and the time-dependent solution. 
The steady-state solution is transformed to the pulsar XLF. 
We showed that this XLF has a broken power-law shape, reflecting the complex behaviour of the efficiency, 
with the high luminosity cut-off which location and shape are determined by the parameters of the birth period and magnetic field 
distributions. 
The location of the cutoff mainly depends on the mean birth period. 
For short enough birth periods, the cutoff may lie above $10^{39}\ergs$. 
Therefore, the existence of  luminous pulsars is possible. 

The time-dependent solution tells us about the evolution of the distribution functions of the pulsars. 
We have shown that at large ages the period distribution becomes a delta-function-like  peaking at $\sqrt{\alpha t}B$.   
The time-dependent luminosity distribution is more complicated due to the complexity of the luminosity-period relation. 
It can be multimodal with different modes related to the different regimes of the efficiency of conversion of the rotation energy 
losses to the X-ray radiation. As the age of the pulsars increases, the luminosity distribution becomes more symmetric. 

We found constraints on the parameters of the birth period distribution 
using the observed XLF of the sources in the nearby galaxies obtained by \citet{mineo12}. 
We found that the mean birth period cannot be shorter than 10--30~ms, depending on the width of the distribution.  
Therefore, the parameters derived by \citet{arz02} lie in the forbidden region for the typically assumed pulsar production rates.  
Accounting for the recent findings of \citet{popov12} the parameters obtained by \citet{kaspi06} are the most reliable. 

We discussed the influence of the beaming and the maximal efficiency on the luminosity function.   
For our calculations we assumed conservatively $b>0.3$, 
but the results can be easily scaled to the different values of the beaming. 
The number of the observed pulsars and their contribution to the ULX population  depend linearly  on $b$. 
The influence of the maximal efficiency is more complex, because it affects only the high luminosity tail of the pulsar XLF. 
For $\etam = 0.3$ the allowed parameter space of the birth period distribution expands towards the shorter birth periods. 
The fraction of pulsars in the observed XLF of \citet{mineo12} would be smaller for smaller values of the efficiency and 
it strongly depends on the  luminosity above which this fraction is computed.  
We showed that for broad initial period distributions, the pulsar fraction is a weak function of $\etam$. 

We have also obtained constraints on the period distribution by applying the method proposed by \citet{perna08}. 
We derived the luminosity function of core-collapse SNe, using published X-ray light curves and compared it
to the time-dependent luminosity function for pulsars.  
We found that the observed luminosities of the SNe are consistent with the mean birth period of $p_0 \gtrsim$0.015--1~s, 
depending on the width of the distribution, maximum efficiency and the beaming factor. 
These constraints are in agreement with those derived by \citet{perna08}. 
 
We estimated a possible fraction of the pulsars in the whole population of ULX, using the observed XLF from \citet{mineo12}. 
For  the  models considered in the previous studies of pulsar populations, the predicted fraction of 
luminous pulsars can be in excess of 3 per cent for the sources with luminosities greater than $10^{39}\ergs$. 
At this moment, about 500 ULXs have been discovered \citep{walton11,Swartz2011,feng_soria11}
and we expect that  at least $\sim 15$ of those should be associated with the rotation-powered pulsars.
The models predict the pulsar fraction above $10^{40}\ergs$ at the level of 1--40 per cent. 

Therefore, we might potentially observe bright pulsars as ULXs in galaxies with high SFR. 
These pulsars should have almost constant luminosity during the first hundred years after their birth,  
but there may exist some  variability on the timescale of $\sim$10~yr related to the interaction of the expanding 
SN remnant shell with the surrounding media.

\section*{Acknowledgments}
The research was supported by the Academy of Finland grant 127512.  


\label{lastpage}

\end{document}